\documentclass[cernpreprint, UKenglish, texmf, orcidlogo]{atlasdoc}
\usepackage{atlaspackage}
\usepackage{atlasbiblatex}

\usepackage{atlasphysics}
\graphicspath{{logos/}{figures/}}

\usepackage{subfig}
\usepackage{floatrow}
\AtlasTitle{Total cost of ownership and evaluation of Google cloud resources for the ATLAS experiment at the LHC}

\AtlasAbstract{%
The ATLAS Google Project was established as part of an ongoing evaluation of the use of commercial clouds by the ATLAS Collaboration, in anticipation of the potential future adoption of such resources by WLCG grid sites to fulfil or complement their computing pledges.
Seamless integration of Google cloud resources into the worldwide ATLAS distributed computing infrastructure was achieved at large scale and for an extended period of time, and hence cloud resources are shown to be an effective mechanism to provide additional, flexible computing capacity to ATLAS.
For the first time a total cost of ownership analysis has been performed, to identify the dominant cost drivers and explore effective mechanisms for cost control.
Network usage significantly impacts the costs of certain ATLAS workflows, underscoring the importance of implementing such mechanisms.
Resource bursting has been successfully demonstrated, whilst exposing the true cost of this type of activity.
A follow--up to the project is underway to investigate methods for improving the integration of cloud resources in data--intensive distributed computing environments and reducing costs related to network connectivity, which represents the primary expense when extensively utilising cloud resources.
}

\author{The ATLAS Collaboration}

\AtlasRefCode{SOFT-2023-02}

\PreprintIdNumber{CERN-EP-2024-124}

\AtlasJournal{Comput. Soft. Big Sci.}
\AtlasJournalRef{Comput. Soft. Big Sci. 9 (2025) 2}
\AtlasDOI{10.1007/s41781-024-00128-x}

\hypersetup{pdftitle={ATLAS document},pdfauthor={The ATLAS Collaboration}}

\begin{document}

\maketitle

\tableofcontents

\section*{Executive Summary}
\label{sec:exec-summary}

The ATLAS Google Project was established to continue an evaluation of commercial clouds, in anticipation of the potential future adoption of such resources by WLCG grid sites to fulfil or complement their computing pledges to ATLAS.
Cost estimates of commercial cloud resources have been done within ATLAS before, but this is the first time a total cost of ownership evaluation was performed for a long--term 15 month period.
Whilst commercial cloud pricing structures are usually fine grained, like most clients ATLAS has negotiated a flat rate subscription agreement with Google for the duration of the project.
Therefore, the method employed here is to analyse the relative contributions of various components within the cloud service to the total cost of ownership, including compute, storage, and network, for different ATLAS workflows and under different operating conditions, and identify the dominant cost drivers.

A substantial amount of technical development work was dedicated to the seamless integration of cloud resources into the ATLAS distributed computing infrastructure, employing the same software stack, primarily through cloud--native interfaces like Kubernetes and signed HTTP URLs following a cloud signature standard like S3v4.
Care was taken to avoid vendor--specific choices and technology to mitigate the risk of potential price volatility in scenarios where one resource or service provider has a monopoly.
The ATLAS Google Project has been key to test and validate at scale. Once these interfaces were established, the operation of an ATLAS site in the cloud has proven to be very effective, unlocking capabilities that would be either unfeasible or considerably time--consuming in an on--premises setting.
For example, it has enabled rapid scaling of the number of processing jobs within a few hours and the deployment of non--x86 architecture resources on a large scale, tasks that could ordinarily take months to accomplish.

From the technical perspective, the project was a success, demonstrating that an ATLAS site can be deployed in a commercial cloud at a very large scale and in a very effective manner, requiring little additional operational effort.
No significant technical issue was discovered to prevent the experiment routinely employing such resources in the future, and the existing workflow and data management services and software packages used by ATLAS are found to be adequate for effectively managing the operation of a significant amount of resources in the Google Cloud Platform.

Over the course of the project, running jobs on the ATLAS Google site may be broken down into several phases.
The initial phase, which is described in some detail, was a process of familiarisation, learning how the site behaves, and applying necessary changes to bring the respective costs of the different services under control.
This was followed by an extended period of running individual ATLAS workflows.
Bursting to a much greater number of cores was then examined, followed by the final phase, where for the last few months of the project both the CPU and storage were expanded to the size of a typical large ATLAS site.

This project has shown that commercial cloud computing is an effective technical mechanism for ATLAS for providing additional CPU resources at the level of a large WLCG site.
Resource bursting was successfully demonstrated, where available CPU resources can be increased by 100,000 additional cores within an hour and with no additional operational overhead.
The utilisation of cloud--based storage was also demonstrated, and the impact of network costs evaluated.
Network egress costs can be very high and currently dominate the overall cost depending on the workloads run at the cloud site.
This study has quantified this effect, whilst also confirming that the ATLAS data management software, Rucio, includes features that allow network traffic and derived costs to be effectively controlled.
The ATLAS computing model currently relies on a plentiful network connectivity between all sites, but it might evolve to reduce data traffic if the actual network costs were exposed.

By leveraging the Google Cloud Subscription Agreement pricing model, ATLAS has effectively harnessed between three and four times the resources compared to what the same investment would deliver for the list--price.
This underscores the vital importance of establishing such agreements with cloud providers, which serve as essential tools for accessing significant volume discounts and ensuring cost predictability, while retaining the flexibility and scalability advantages inherent to cloud services.
In essence, these agreements are not merely advantageous but rather a prerequisite for enabling large--scale cloud deployments.
As such, the list--price should therefore be seen as the upper limit of the actual price paid for large scale cloud services.

Whilst many valuable insights into the use of commercial clouds have been provided, it is still possible to further develop the ATLAS workload and data management software stack in order to integrate commercial clouds in the most efficient way.
Key areas for future work include evaluating how the private cloud and LHC research network infrastructure can be interconnected and how the orchestration of data and workflows can provide maximal gains in the performance and flexibility of the computing model with minimal additional cost.
The full implementation of these developments is important to enable exploring a potential evolution of the ATLAS computing model that tackles the issue of network costs.
An extension of the project would also enable the current, wide ranging R\&D programme to continue, which exploits the elastic availability of special resources such as GPUs and alternative architectures such as ARM, otherwise not readily available to ATLAS, enabling fast validation and benchmarking of new architectures without the need to make upfront investment in hardware.

\section{Introduction}
\label{sec:introduction}

The ATLAS experiment~\cite{PERF-2007-01} at the LHC~\cite{Evans_2008} employs distributed computing resources of up to one million cores of computing, 350~PB of disk and over 500~PB of tape storage.
These resources comprise the Tier--0 at CERN, the Tier--1 and Tier--2 Worldwide LHC Computing Grid (WLCG) sites~\cite{CERN-LHCC-2005-024,CERN-LHCC-2014-014}, opportunistic resources at High Performance Computing (HPC) sites and cloud computing providers, as well as volunteer computing.
In recent years such non--WLCG resources, particularly from HPC sites, have made increasingly significant contributions to ATLAS computing, a trend that is expected to continue in the future.

The use of commercial clouds has been investigated by all LHC experiments~\cite{García_2014,panitkin2015look,Holzman2017,Loncar:2023mog}, and was recently reviewed by the US ATLAS and CMS communities as part of a blueprint publication~\cite{Megino:2023zch}.
ATLAS has examined the viability of both Google and Amazon cloud resources~\cite{megino2021seamless}, including the integration of the two core components of the distributed computing environment: the workload management system PanDA~\cite{megino2017panda} and the data management system Rucio~\cite{barisits2019rucio}.
These initial studies showed that such resources can be adopted for both the ATLAS production and user analysis workflows, albeit at a limited capacity, as well as demonstrating the potential for future R\&D~\cite{megino2021seamless}.

As the start of Run 4 and the HL--LHC~\cite{Apollinari:2015bam} era approaches, the need for additional CPU and especially disk resources continues to grow, to satisfy the computing requirements of the experiments~\cite{ATLAS-HL-LHC-CDR,ATLAS-HL-LHC-Roadmap}.
At the same time, WLCG sites are increasingly exploring methods to utilise new resources, aiming to fulfil their pledges in the most cost--effective and energy--efficient manner while also making emerging technologies accessible to users.
With this in mind, the ATLAS Google Project (AGP) was established to continue an evaluation of commercial clouds as a resource for ATLAS.

Whilst basic cost estimates of commercial cloud resources have been done before~\cite{Holzman2017,megino2021seamless}, one of the primary goals of the AGP is to develop a Total Cost of Ownership (TCO) model for commercial clouds, which is made possible by the highly granular pricing information available from the cloud provider.
By comparison, other types of computing resources employed by ATLAS, namely the grid, HPCs, clouds, and volunteer computing, incur different effective costs to the experiment, which are often difficult to evaluate and compare since funding methodologies vary by resource type, by country, and even between different funding agencies within the same country.
Additionally, there are local arrangements at the sites or savings, for example where some sites, typically at universities, do not directly pay electricity or WAN access costs.
It is therefore important to note that this TCO evaluation is not attempting to directly compare the cost of running ATLAS jobs in the Google Cloud with the cost of running an ATLAS grid site.
A TCO comparison of data centre colocation and commercial cloud solutions has been previously performed in the context of expanding computing resources at CERN~\cite{cernTCO}, using input from the Helix Nebula Science Cloud project~\cite{helixNebulaCloud}.

The structure of this paper is the following.
Section~\ref{sec:cost-modelling} delves into the critical elements of commercial cloud cost models relevant to this study, examining the subscription agreement model employed by ATLAS and outlining the methodology and scope of the TCO analysis.
Section~\ref{sec:site-integration} briefly outlines the technical aspects of integrating the Google site and Section~\ref{sec:running-the-site} offers a detailed account of the various tests conducted at different phases and the resulting conclusions.
Section~\ref{sec:network-costs} more closely examines the role of network traffic in overall costs and underscores the rationale for considering dedicated links.
Section~\ref{sec:feedback} presents feedback received from a number of system administrators regarding the adoption of commercial clouds.
Finally, Section~\ref{sec:future} provides an outline of potential future working directions and a summary is presented in Section~\ref{sec:summary}.

\section{Commercial cloud cost modelling}
\label{sec:cost-modelling}

In contrast to other computing resources employed by ATLAS, including those deployed via the WLCG grid, commercial cloud resources have well defined pricing for computing time and services used, for example at Amazon Web Services (AWS)~\cite{awsPricing} or on the Google Cloud Platform (GCP)~\cite{gcpPricing}.
While each cloud may differ in their specific pricing structure, the cost of the individual services used by the client is usually published in an itemised and highly granular way, making it easier to develop future cost models for commercial clouds.
In this way, a client can choose from a menu of services, which are charged or billed directly.

While the \emph{list--price--based} cost model is suitable for ad hoc or specialised short--term resource needs, it is likely to prove more costly for consistent, long--term usage of resources compared to grid--based, general--purpose offline computing.
However, most commercial cloud providers offer discounts and credits for large--scale users, which has been previously explored by ATLAS with varying degrees of success.
Such discounts and credits can significantly reduce the cost compared to the list--price.
Funding agencies can often negotiate even better deals.
The list--price--based cost model therefore provides a maximum ceiling for a TCO, but rarely reflects the actual price paid for large scale cloud services.

The primary obstacle to using commercial clouds is usually egress costs, which are network costs incurred when data is exported from the cloud resource, and may be prohibitively high for the distributed computing workflows used by ATLAS.
These workflows often exploit the high interconnectivity of the grid sites and incorporate many data transfers, as data hosted by one site is transferred to be used elsewhere.
This high network interconnectivity may incur significant expense, which goes on top of the site budgets, as discussed in Section~\ref{sec:dedicated-networks}.
Egress costs associated with cloud resources were a concern expressed by several Tier--1 site administrators when interviewed for this report (see Section~\ref{sec:feedback}).

Whilst discounts, credits and subscription plans may increase the complexity in determining future costs, they also provide mechanisms to reduce the total cost of using commercial clouds for ATLAS.
It should however still be noted that even with deep discounts and credits, continuous monitoring of expenditure must be done, and as automated as possible, especially for large--scale use.

\subsection{Subscription model employed by the ATLAS Google Project}
\label{sec:subscription-model}

For this latest phase of the AGP, a \emph{Google Cloud Service Agreement for Public Sector} contract was negotiated with Google to explore more solid ideas about employing Google as an ATLAS production site.
This contract, which uses the latest pricing model, was negotiated via US national laboratories and is based on an initial assumption of an average of 7,000 compute cores together with up to 7~PB of storage, and an estimated egress of no more than 0.7~PB per month.
As is often the case with commercial cloud resources, there is no charge for data ingress, that is for uploading ATLAS data to the Google Cloud Platform.
The contract ran for 15 months, from July 2022 to October 2023, at a flat rate cost of \$56,630.54 per month, which resulted in a subscription price of around \$1900 per day.

The subscription model does not put a limit on usage once active, and ATLAS can use the available resources at any time at any scale.
For example, for one month the disk usage could be 10~PB, and the next month 1~PB, or for one month ATLAS may use 3,000 cores, and the next month use 20,000 cores.
This resource elasticity may be very useful for short duration data processing campaigns and is explored in dedicated tests.

The TCO for the subscription model is the total price of the contract for 15 months, \$849,458, although there is a clear caveat to be considered, namely that the negotiation for any subsequent contract will likely examine the actual average usage during the previous contract.
If a significantly higher usage is observed than the initial estimate, it may be the case that the monthly price may be higher in any subsequent contract.
A key part of the TCO evaluation is therefore to use the detailed usage breakdown provided by the GCP accounting, without applying the subscription model, to understand which site configurations produce the most significant increase to the total cost according to the list--price, and which particular services drive this increase.

\subsection{Total cost of ownership methodology}
\label{sec:tco-methodology}

As previously mentioned, this TCO evaluation is not trying to directly compare the cost of running an ATLAS grid site with the costs associated with running ATLAS jobs in the Google Cloud.
This is primarily for two reasons, as described above.
Firstly, a significant cost variation exists already among different resources, and among different grid sites, which is at least in part attributable to the variation in service costs, as well as regional variation in labour costs for site administrators.
Secondly, relying solely on absolute numbers for Google Cloud costs may not provide meaningful insights, due to the substantial influence of individual agreements and contracts, which include volume discounts that vary case by case and over time.

The focus of this TCO is therefore rather on gaining a comprehensive understanding of how to make the best use of such a resource, including how the configuration may differ from a standard grid site.
To accomplish this, the relative contributions of various components within the cloud service to the TCO are analysed, including compute, storage, and network, across different operating models.
By identifying the dominant cost drivers and exploring effective cost control mechanisms, it is possible to optimise resource allocation and management to maximise cost efficiency.

In addition to hardware considerations, the invaluable contribution of dedicated personnel at the grid sites should not be overlooked.
These individuals not only fulfil the role of system administrators but also play a vital part in maintaining the middleware and the distributed computing infrastructure of the experiment, and in some cases contribute to federated support services, including user--support, and areas such as R\&D, outreach and education.
Their expertise and involvement are crucial to sustain a high efficiency and effectiveness in the operation of the ATLAS computing infrastructure, more so if sites opt for using cloud computing resources at scale.

\section{ATLAS Google site integration}
\label{sec:site-integration}

The ATLAS data are distributed worldwide across data centres or sites, organised into Tiers with varying capacities and responsibilities under the umbrella of the WLCG~\cite{CERN-LHCC-2005-024,CERN-LHCC-2014-014}.
The single Tier--0 centre is CERN, and there are ten Tier--1 sites, connected via dedicated National Research and Education Networks (NRENs) typically with between 10 and 100~Gbps, using the LHCOPN and LHCONE overlays~\cite{chep23networks}.
The Tier--1 sites provide both disk and tape storage and function as the perpetual archival of the collision data.
Around 50 Tier--2 sites, typically hosted by national universities and laboratories, provide disk storage that is used for data processing and user analysis.

When setting up an ATLAS site using GCP resources the objective is to fully support all ATLAS production workflows.
This includes not only processing the RAW collision data into reconstructed data (Analysis Object Data, AOD), but also running all components of the multi--step Monte Carlo (MC) production workflow, such as Event Generation (EVNT), Full or Fast Simulation which produces simulated detector interaction data (HITS), simulated detector output data (Raw Data Object, RDO) and reconstructed simulated data (AOD), as well as creating data and MC derived formats (DAOD) for input to analysis (a process referred to as ``Group Production'').
Additional formats such as Derived Event Summary Data (DESD), which are tailored to the specific needs of various subdetector and object reconstruction and identification performance groups, may also be produced in data processing campaigns.
The site is also expected to handle user workloads, which can encompass a diverse range of demands.
Further details on ATLAS production workflows and data formats can be found elsewhere~\cite{run3paper}.

In the ATLAS grid site setup, production tasks running ATLAS jobs at various locations consolidate their outputs at a designated site known as the nucleus.
The nucleus is chosen during task definition and can be either a Tier--1 site or a substantial and reliable Tier--2 site.
Considering its size and expected performance, this responsibility could also be anticipated from the ATLAS Google site and is also examined here.

The integration of compute at cloud based sites with ATLAS distributed computing relies on Kubernetes~\cite{k8s}, where the resource--facing component of PanDA, Harvester~\cite{maeno2019harvester}, utilises the native job controller of the Kubernetes clusters for submitting batch jobs to the PanDA queue associated with the site~\cite{atlasKubernetes,chep23clouds}.
ATLAS software is provided in the same way as at grid sites via CVMFS~\cite{blomer_2020_4114078}.

The access to Google storage was configured as a standard \emph{Rucio Storage Element} (RSE), as used by other WLCG storage systems abstracted via standard HTTP/WebDAV using authorisation tokens based on the S3v4 format~\cite{s3v4}.
At the lowest layer in the stack, the Davix~\cite{davix} library implements HTTP/WebDAV access to all storage systems used in the WLCG, and already supports chunked transport required by object stores, which are used by ATLAS for cloud based sites~\cite{chep23rucio}.
Further significant development was then required to make access to Google storage work, as described in the following.

The Rucio server was extended to allow only specific accounts to access cloud storage and to generate cloud storage tokens ad--hoc when listing replicas, a functionality that is needed both by interactive command line interface users as well as production and analysis jobs.
It was also necessary to extend the functionality of the Rucio clients to allow seamless upload/download for object stores, as object stores prohibit several useful functions from the HTTP/WebDAV standard that Rucio uses to ensure safe uploads such as checksum verification.
The CERN File Transfer Service (FTS)~\cite{fts} was also extended to dynamically generate authorisation tokens based on the S3v4 format when enacting a transfer.

Making the Google storage available within the WLCG infrastructure also required a creative authentication solution, as Google is not part of the Interoperable Global Trust Federation (IGTF)~\cite{igtf}.
To do this, a dedicated load--balancer was set up on the Google Cloud side with a fake hostname in the CERN DNS server such that a CERN--based X.509 host certificate could be issued and uploaded.
This load--balancer was configured based on path--based regular expressions, which allowed two RSEs to be deployed: a DATADISK to store production input and output data and a SCRATCHDISK to temporarily store analysis job outputs.
With this setup, the Google storage could be globally integrated into the ATLAS distributed computing infrastructure like any other WLCG site, and could be used by any account with the appropriate permissions.

Due to the proximity to CERN and the low carbon--intensive energy usage in the region, the Google \emph{europe--west1} region in Belgium was chosen to host the primary ATLAS Google site, although during investigations into the network connectivity a second site in the US Google \emph{us--east4} region in Virginia was also employed (see Section~\ref{sec:network-costs}).
The number of running single--core slots (``job slots'') at the ATLAS Google site can be manually configured, and is typically set to either 5,000 or 10,000, although additional set--ups such as those used for the evaluation of larger scale “bursts” were also deployed, as described in the Section~\ref{sec:bursting}.
CPU cores are provided as “spot instances”~\cite{gcpSpotvms}, so that allocated resources may be preempted at any time depending on current situation at the cloud, but have the advantage of costing significantly less in real terms.
The storage limit at the ATLAS Google site is initially configured to be between 2 and 5~PB.

Thanks to the development of support for cloud--native interfaces, the deployment of an ATLAS site in the Google Cloud capable of scaling to tens of thousands of CPUs and several petabytes of disk could be accomplished within a matter of weeks.
Moreover, the operation and maintenance of the site has required only a fraction of a full--time equivalent (FTE) staff with expertise in cloud administration.
This achievement paves the way for exploring cost--benefit scenarios in which the agility and scalability of cloud computing can be harnessed to accelerate the ATLAS science programme.

\section{Running the ATLAS Google site}
\label{sec:running-the-site}
\subsection{Initial phase}
\label{sec:initial-phase}

\begin{figure}
\centering
\sidesubfloat[]{\label{fig:firstSixMonthsA}
\includegraphics[width=0.9\textwidth]{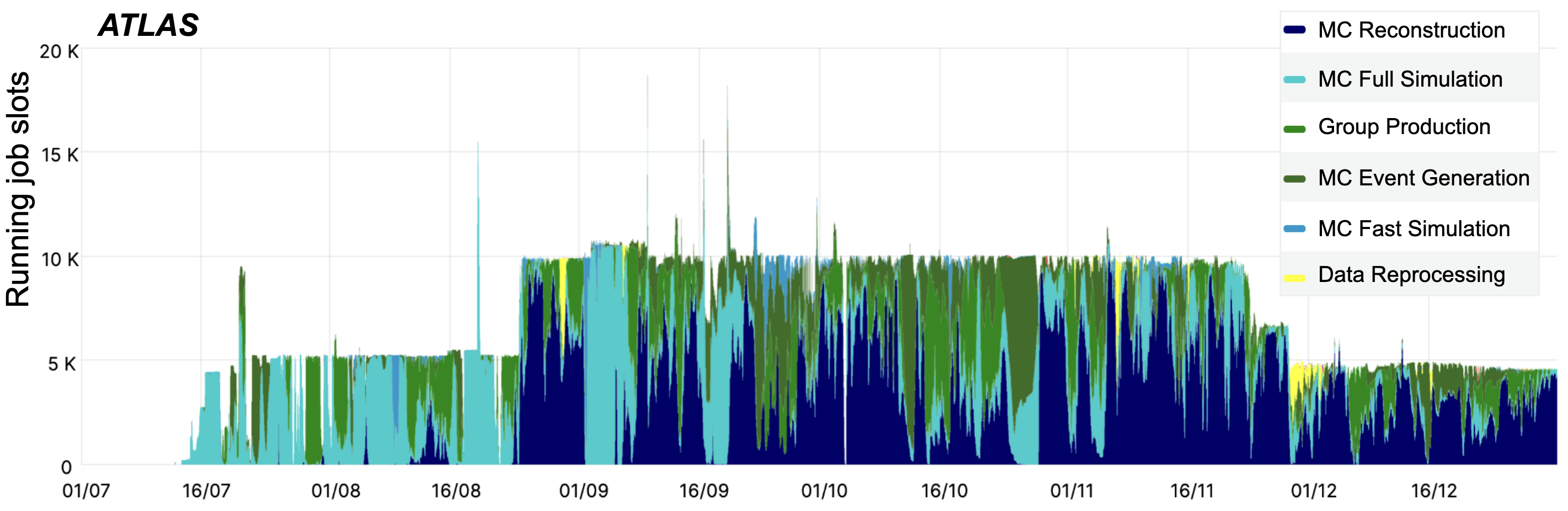}
}
\vspace{0.3cm}\\
\sidesubfloat[]{\label{fig:firstSixMonthsB}
\includegraphics[width=0.9\textwidth]{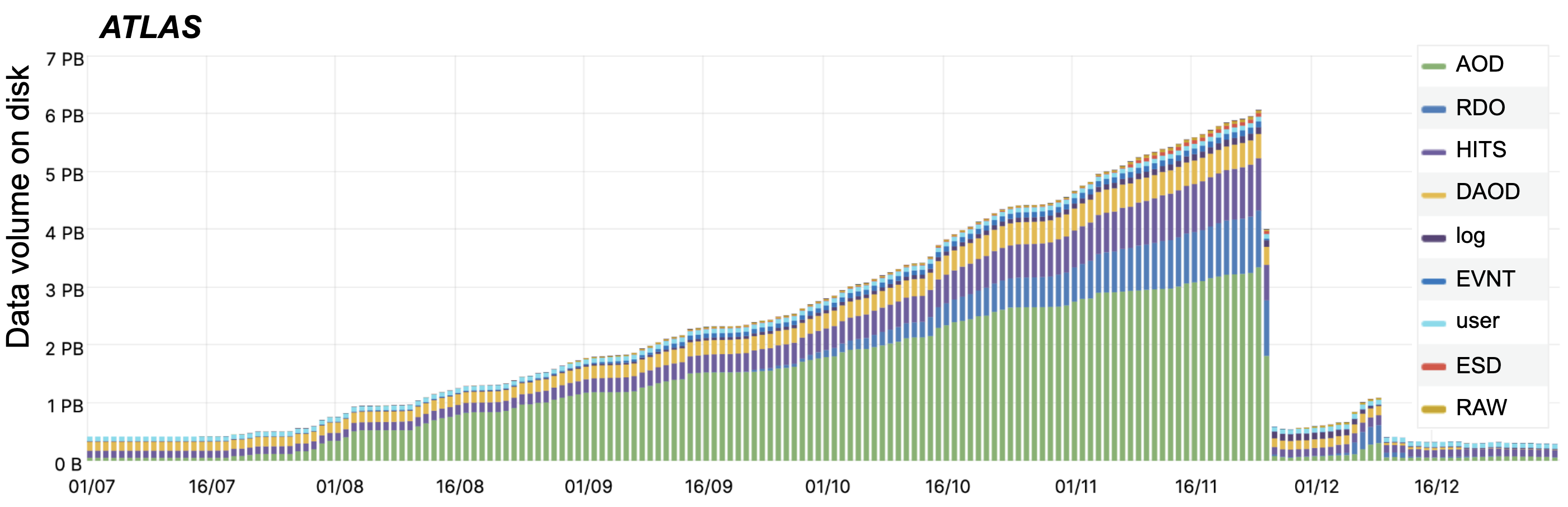}
}
\vspace{0.3cm}\\
\sidesubfloat[]{\label{fig:firstSixMonthsC}
\includegraphics[width=0.9\textwidth]{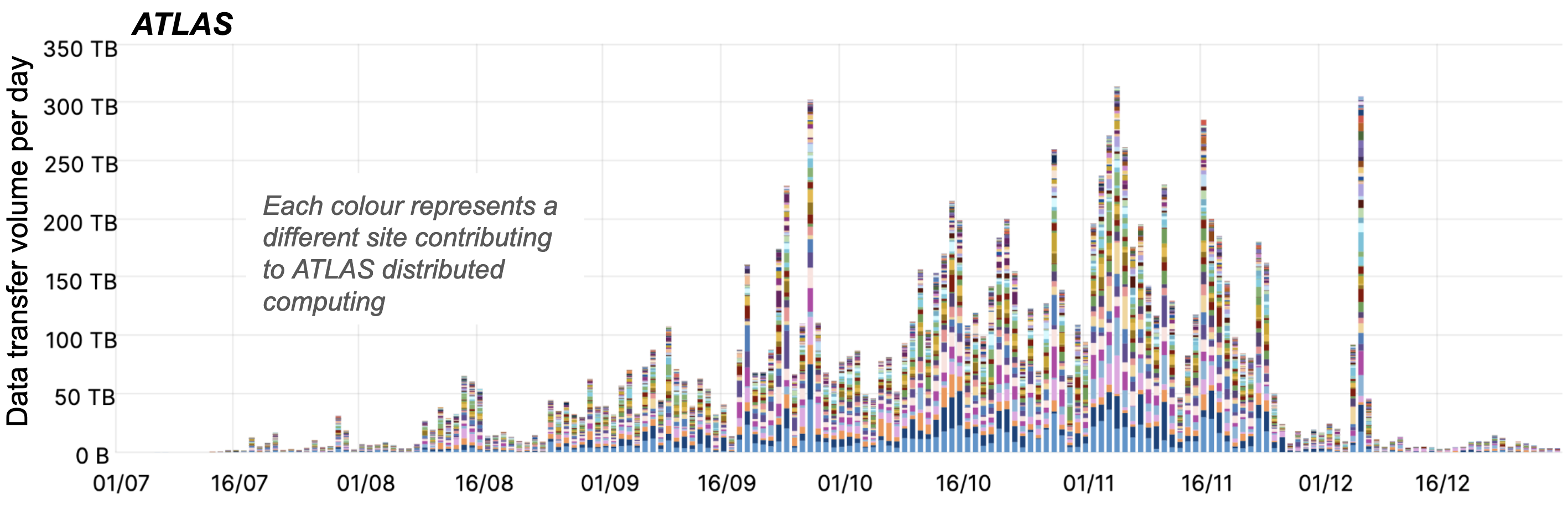}
}
\vspace{0.3cm}\\
\sidesubfloat[]{\label{fig:firstSixMonthsD}
\includegraphics[width=0.9\textwidth]{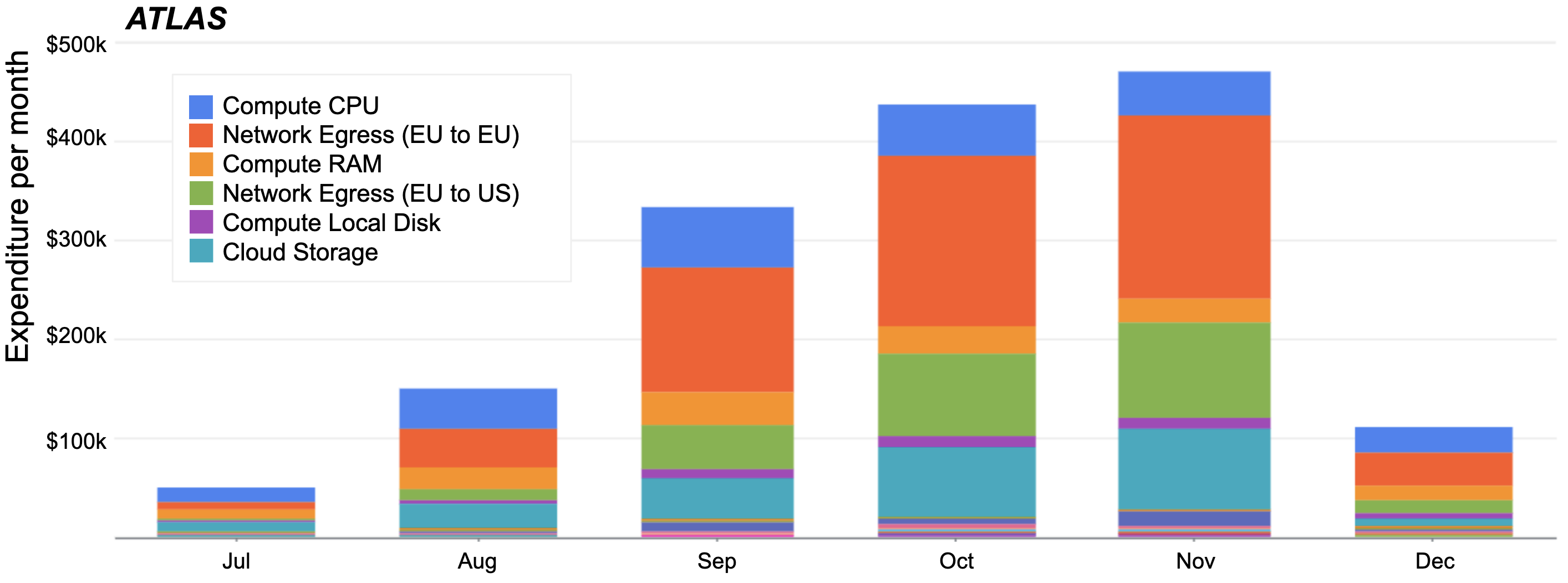}
}
\caption{Monitoring plots for the first six months of running at the ATLAS Google site, from July to December 2022. (a) The number of running jobs at the Google site. (b) The accumulated data at the Google RSE split into different formats, the main ones being AOD (green), RDO (blue), HITS (purple) and DAOD (yellow). (c) The daily egress traffic out of the ATLAS Google site, split into the various destination sites. (d) The monthly list--price cost per service from the Google billing console, where the six main components are shown in the legend.}
\label{fig:firstSixMonths}
\end{figure}

The ATLAS Google site was initially configured with a PanDA queue able to run up to 5,000 CPU slots, together with a single RSE where data files could be stored.
The PanDA queue could be configured to accept certain types of jobs and the initial goal was to try to increase the number of different workloads and eventually test all the job workflows at the site.
Figure~\ref{fig:firstSixMonths} displays various metrics of the ATLAS Google site during the first six months of running.

The number of running jobs is shown in Figure~\ref{fig:firstSixMonthsA}. Aside from alternating the number of CPU slots between five and ten thousand, the configuration of the site was essentially unchanged during this period.
The number of different types of job running in the PanDA queue was progressively increased by changing the brokerage decisions to adjust the job mix of the PanDA queue.
The thin spikes that can be seen in the number of running jobs are due to short--term site configuration changes.

In the three months from August 24th until November 24th 2022, the overall job mix that was run corresponded to approximately 30\% MC Event Generation, 30\% MC Full Simulation, 30\% MC Reconstruction and 10\% Group Production, which is a typical job mix seen on a standard grid site.
Analysis workloads were not tested in this period as many changes needed to be made to the ATLAS middleware.

As jobs continued to run at the site, the accumulated data from the various production steps steadily increased at the Google RSE at a rate of approximately 50~TB per day, until it reached 6~PB on November 24th as can be seen in Figure~\ref{fig:firstSixMonthsB}.
At the same time, the availability of these data generated an increasing number of accesses from other ATLAS sites.
The egress network traffic from the ATLAS Google site due to production and analysis jobs elsewhere using data at Google as input ramped up from an average of about 20~TB per day in August to about 130 TB per day in October and November.
In November, there were periods with egress network traffic over 200~TB per day for several days in a row, as can be
seen in Figure~\ref{fig:firstSixMonthsC}.
It is worth noting that this traffic is above the average seen at ATLAS sites, relative to the stored data volume.
The ATLAS Google site was generating egress traffic at the level of 4~PB per month in November whilst hosting 5--6~PB data, which is significantly more than the initial estimate of 0.7~PB.
By comparison, the MWT2 grid site in the US also generates an average of 4~PB per month of egress traffic, whilst hosting more than 15~PB of data.

Figure~\ref{fig:firstSixMonthsD} shows the breakdown of the Google list--price cost per service, where the various contributions from compute, storage and egress are presented.
It can clearly be seen that the costs of egress traffic and storage can quickly become dominant in cloud resources.
To contain these costs, on November 25th the configuration of the ATLAS Google site RSE was changed in Rucio by enabling greedy deletion, so that data were deleted as soon as the corresponding replication rules had expired~\cite{rucioReplicas}.
This had the immediate effect that the 5.5~PB of accumulated cached data were quickly removed, and from then on new temporary data would stay on the RSE for only one or two weeks. This drastically reduced the egress traffic, since before the change there was significant egress as the cached data were being transferred multiple times to other sites as job inputs.

After this change, the remaining egress was mainly due to job output being sent to the task nucleus. Another change was then applied to the Google RSE on December 8th that set the distance~\cite{rucioRses} of the RSE to any other ATLAS site to a very large value.
This had the effect of completely eliminating the remaining egress traffic due to job inputs then preferentially being read from other sites.
The number of CPU slots at the ATLAS Google site was also reduced back to five thousand.
The effect of these changes can be seen in the distributions in Figure~\ref{fig:firstSixMonths}.

With this new configuration, the stored data at the Google RSE stayed around 300 TB and egress traffic fell below 5--10~TB per day.
The solitary spike in transfer volume in December visible in Figure~\ref{fig:firstSixMonthsC} is from the egress of the outputs of a small data reprocessing running at the ATLAS Google site, visible as the yellow contribution in Figure~\ref{fig:firstSixMonthsA}.

\subsection{First observations on network considerations}
\label{sec:first-network}

Cloud resources have well defined price structures for each resource type, not only compute and storage but also network traffic.
The first two are the usual capacity metrics that are closely accounted for in a distributed infrastructure like the WLCG.
The network capacity is certainly also considered within WLCG, but the planning and provisioning cycle is usually done on a longer time scale.
This is in part because the administration domain for networks is typically broader and often spans national or continental institutions beyond the sites themselves.
This has probably had an effect in the experiment computing models, whose workflows rely to some extent on a plentiful any--to--any connectivity.

The long--term large--scale test of the ATLAS Google site has provided useful insights into the management of the associated network traffic.
Given that the site was newly deployed, it was possible to monitor how the consumption of different types of resources evolved.
CPU usage was constant as expected, as a fixed configuration parameter, whereas storage usage increased at an essentially constant rate.
The steady increase in egress network traffic was correlated with increased use of storage, albeit with a much larger variability.
An interesting outcome of this first period of the ATLAS Google site was that it was possible to reduce the egress network traffic by adjusting a few parameters in Rucio, allowing the cost of cloud resources to be very effectively controlled.

\begin{figure}
\centering
\includegraphics[width=\textwidth]{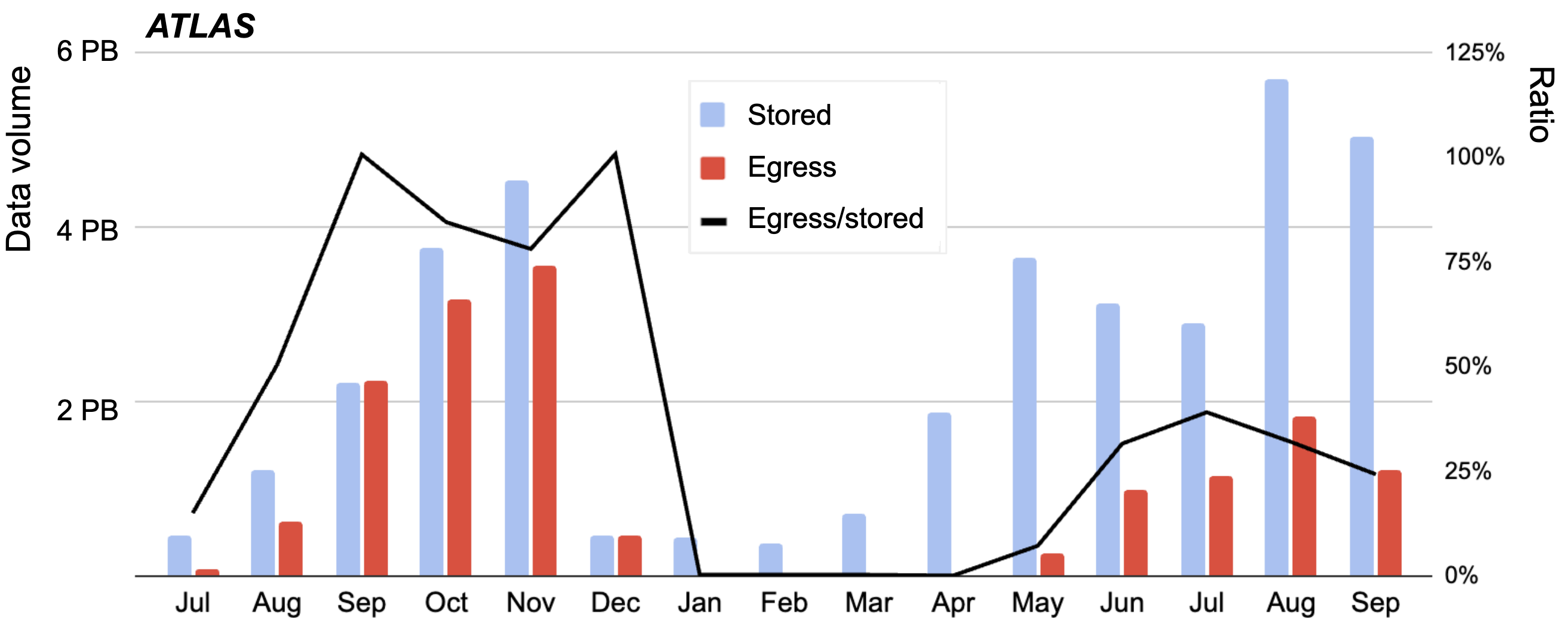}
\caption{Data stored (blue) and egressed for job inputs (red) at the ATLAS Google site per month from July 2022 to September 2023. The ratio is also indicated by the black line.}
\label{fig:dataStoredAndEgressed}
\end{figure}

A key question remains about how useful a grid site is with limited egress network traffic.
Whilst this question is beyond the scope of this report, it is an important topic that deserves dedicated studies in the future in the context of computing model evolution.
The approach taken here is to quantify the tests that were done with the ATLAS Google site.
As previously described, after the first 3--4 months of continuous unattended operations, the site had accumulated 6~PB of data on disk.
The values for average stored data and total egress of data for job inputs are depicted in Figure~\ref{fig:dataStoredAndEgressed}.
Remarkably, between September and December 2022, between 75\% and 100\% of all the data at the site was egressed each month for production or analysis job inputs.
Comparing this with other large ATLAS sites, typically having storage sizes ranging from 4 to 30~PB, this metric falls to between 15\% and 20\%.
Consequently, the significant amount of new data at the ATLAS Google site generated data movement dynamics that significantly deviated from the average.
This is relevant when evaluating a cloud resource, since abnormally high egress traffic levels will have an impact
on costs.

\subsection{Understanding the cost impact of different ATLAS workflows}
\label{sec:different-workflows}

\begin{figure}
\centering
\sidesubfloat[]{\label{fig:singleWorkflowsA}
\includegraphics[width=0.96\textwidth]{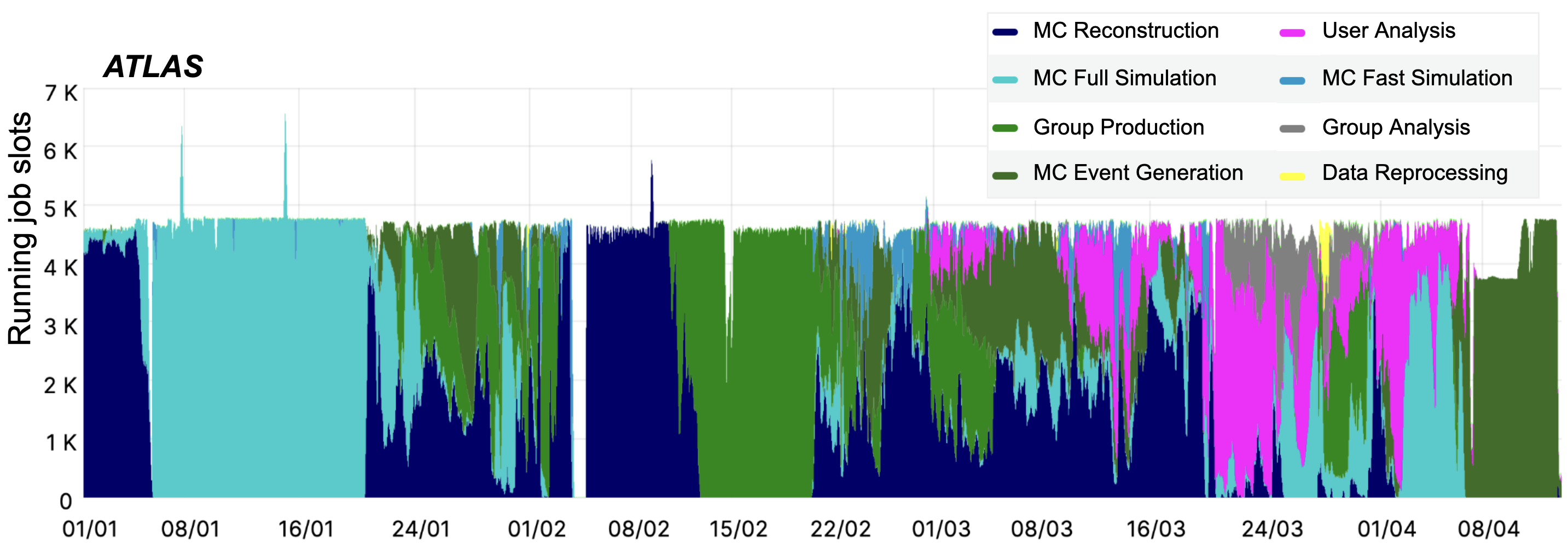}
}
\vspace{0.3cm}\\
\sidesubfloat[]{\label{fig:singleWorkflowsB}
\includegraphics[width=0.96\textwidth]{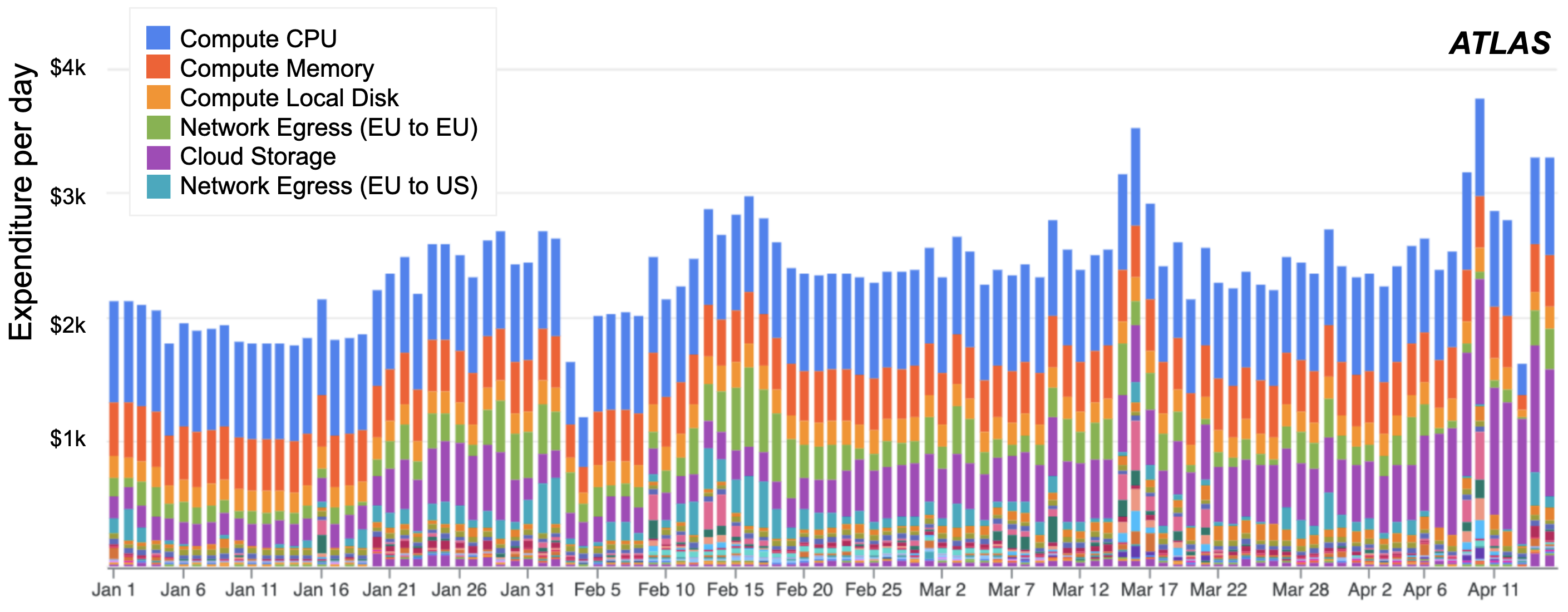}
}
\caption{(a) The variation of workflows running at the ATLAS Google site from January to April 2023 featuring several periods of running with a single workflow. The contribution from User Analysis jobs can be seen from March. (b) The daily list--price cost per service from the Google billing console for the period from January to April. The dominant services are compute CPU/RAM (blue/red), local storage on the worker nodes (orange), cloud storage (purple) and network egress (green and turquoise).}
\label{fig:singleWorkflows}
\end{figure}

Between January and April 2023 the ATLAS Google site ran at an approximately constant CPU capacity of around 5,000 job slots.
The types of jobs allowed to run were controlled through the fairshare policy parameter associated with the PanDA queue.
By adjusting this parameter, four periods of running only one activity were carried out, as shown in Figure~\ref{fig:singleWorkflowsA}: a 14 day period with only MC Simulation jobs in January, a five day period with only MC Reconstruction jobs in early February, a nine day period with only Group Production jobs in mid--February and a six day period of only MC Event Generation jobs at the beginning of April.

The Google billing console provides daily list--price cost information, shown in Figure~\ref{fig:singleWorkflowsB}, which can be used to infer general trends.
It was observed that 80--90\% of the cost is consistently dominated by three services: compute, storage and network egress.
The remainder comes from a combination of infrastructure overhead and costs associated with orthogonal ATLAS R\&D activities at Google.
The compute cost (which has three components associated with it: CPU, RAM and local disk) remains pretty much constant throughout this period, as expected from the fact that the number of slots in the ATLAS Google site was set to 5,000 job slots and not modified.
This cost is consistent with the advertised pricing for the spot \emph{n2--standard--8} instances of around \$0.01 per CPU--hour.
This situation can be compared to the first months of the project, where egress costs dominated, as shown in Figure~\ref{fig:firstSixMonthsD}.

Compute is the dominant cost for these four workflows.
Storage and network costs for a given day are partially but not fully correlated with the activity that is running at that time. If data are not purged from the storage, the storage costs will reflect the accumulation of any past activity.
On the other hand, if there are old data stored at the site that are accessed from outside for any reason, this will generate
egress costs that are not correlated with the activity running at that precise moment.
Still, the single workflow testing periods were of relatively short duration and the Rucio storage configuration meant that no background egress was present, so any correlation observed should be meaningful.

A final, significant single workflow test was done in July 2023, by running a fraction of the reprocessing of the 2022 data on 10,000 job slots at the ATLAS Google site.
About 15\% of the proton--proton collision data recorded in 2022 for physics analysis, comprising around 600M events, were processed into multiple formats for analysis and further, secondary processing between July 11th and 18th.
The data carousel mechanism~\cite{borodin2021atlas} was employed as done for the grid, whereby input files, in this case RAW data, were recalled from ATLAS tape resources and replicated to the Google site, without observing any adverse effects.

\begin{figure}
\centering
\sidesubfloat[]{\label{fig:dataReproA}
\includegraphics[width=0.96\textwidth]{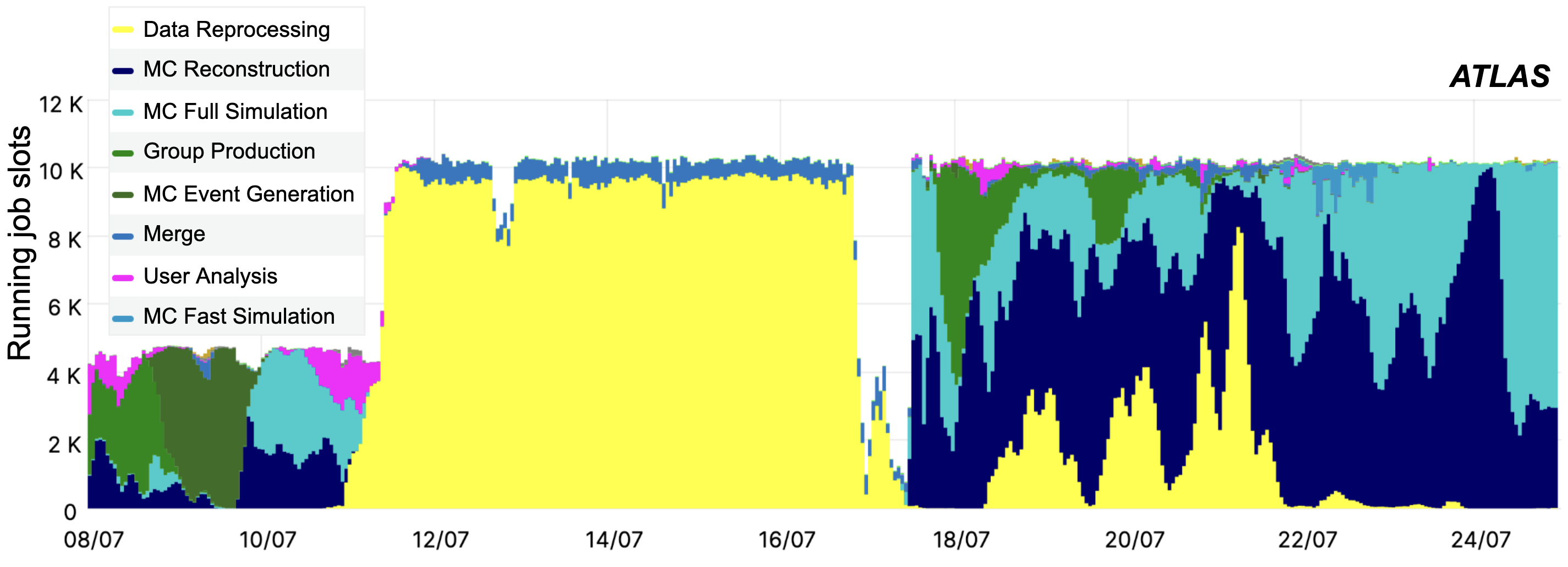}
}
\vspace{0.3cm}\\
\sidesubfloat[]{\label{fig:dataReproB}
\includegraphics[width=0.96\textwidth]{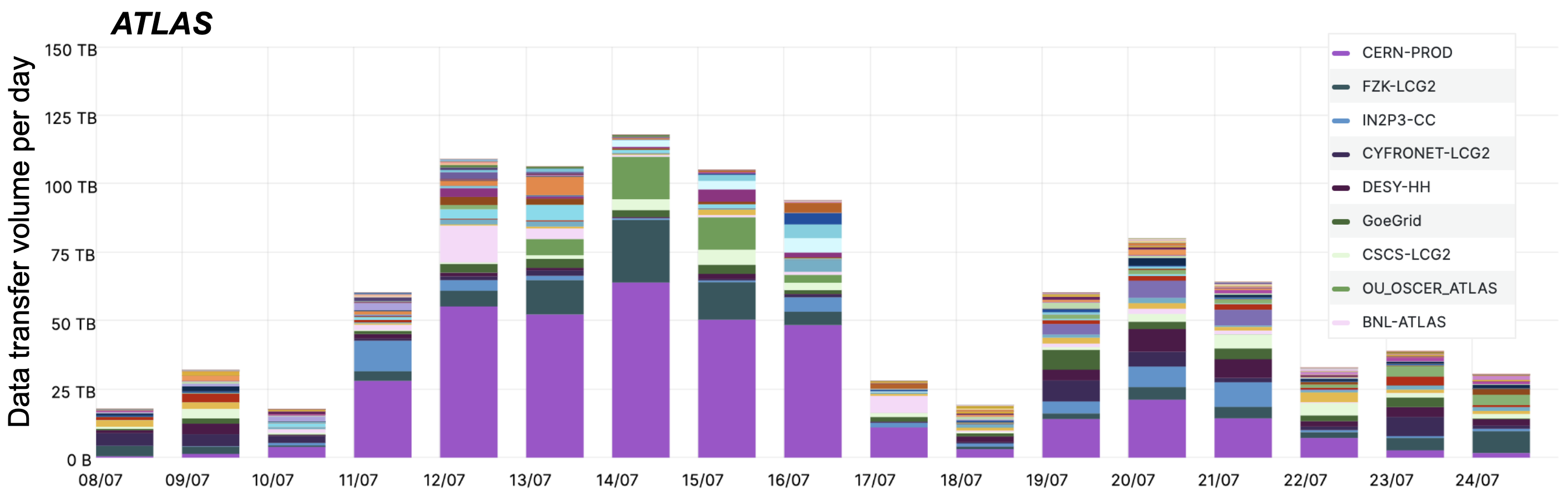}
}
\vspace{0.3cm}\\
\sidesubfloat[]{\label{fig:dataReproC}
\includegraphics[width=0.96\textwidth]{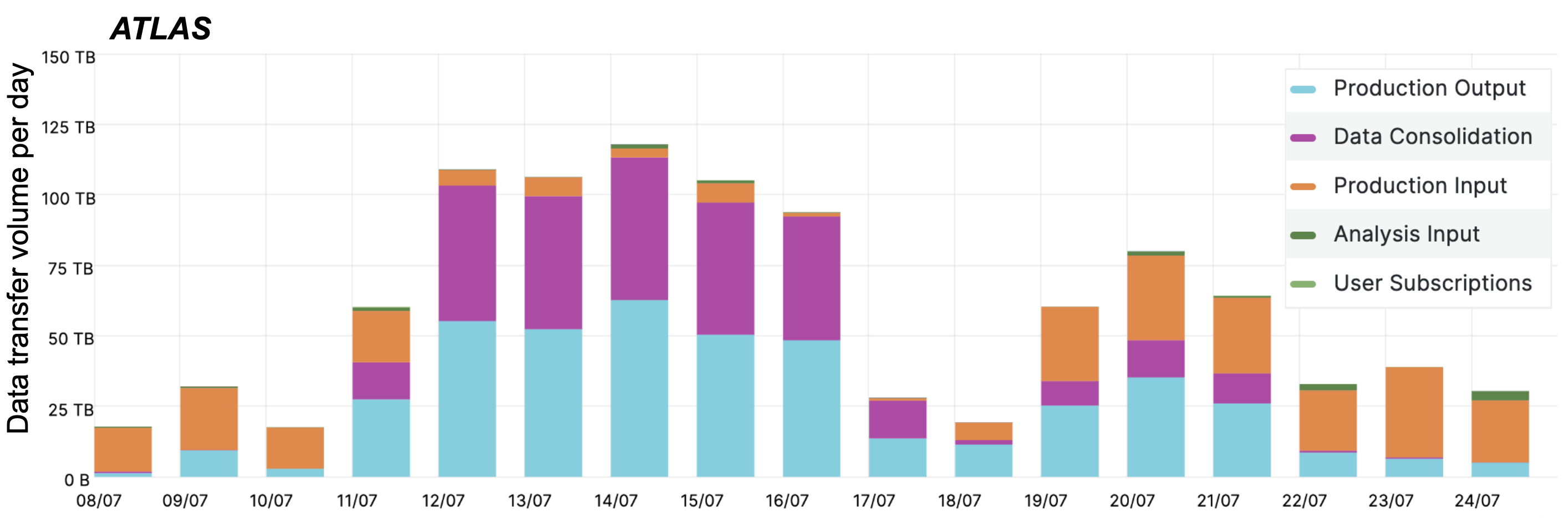}
}
\caption{Distributions covering the data reprocessing campaign performed on the Google ATLAS site. (a) Number of running jobs, where the single job type period from July 11th to July 18th shows the data jobs in yellow, together with a small number of associated merge jobs in blue. (b) The data transfers out of the ATLAS Google site for the same period to different grid sites, where the main contribution in dark purple is the replication of the reprocessing output data to CERN. This is also visible for the data reprocessing jobs after July 18th. (c) The different types of transfers out of the ATLAS Google site for the same period.}
\label{fig:dataRepro}
\end{figure}

Figure~\ref{fig:dataReproA} shows the data reprocessing jobs during the period July 11th to July 18th, as well as a small number of associated output data merging jobs.
Further data reprocessing jobs can be seen in Figure~\ref{fig:dataReproA} after July 18th, which are part of the wider campaign to reprocess the remaining 85\% of the 2022 data on all ATLAS distributed computing sites, including the ATLAS Google site.
Figure~\ref{fig:dataReproB} shows the daily data volume transferred out of the ATLAS Google site to the various grid sites, where it can be seen that during the data reprocessing a total of about 100~TB of data were exported daily.

Figure~\ref{fig:dataReproC} also shows the daily data volume transferred, but now broken down into the different activities.
Production Output (cyan) accounts for approximately half of the egress, which is equivalent to the export to CERN shown as the purple component in Figure~\ref{fig:dataReproB}.
Most of the remaining egress during this period is attributable to Data Consolidation, which is a rebalancing procedure performed on the ATLAS distributed computing system as a whole.
It may be interpreted as other data moved out of the Google storage to make room for the output of the data reprocessing.
The egress from the data reprocessing jobs after July 18th is also visible in Figures~\ref{fig:dataReproB} and~\ref{fig:dataReproC}, albeit at a lower level.

The periods with different types of activities show different trends on the relative cost of the three main services, as shown in Table~\ref{tab:singleWorkflowCosts}.
The average values for the full duration of the project are also displayed.
For the first four columns, a few trends are visible in the numbers in the table.
In the Group Production period, the egress network activity and its associated costs show a clear increase, averaging 21\% over the nine--day testing period.
In the six day MC Event Generation period, the storage is seen to contribute a significant fraction of the total cost at 30\%; however this is from the replication of almost 1~PB of DAOD data sets to the Google site done during the same week.
The most striking observation is that the period of Data Reprocessing activity shows a very different pattern compared to the others, with the network egress cost clearly dominating and averaging 63\% of the total cost.
The relative costs of different workflows are further examined in Section~\ref{sec:network-costs} when discussing networks.

\begin{table}[t]
\centering
\setlength\tabcolsep{5pt}
\renewcommand{\arraystretch}{1.1}
\caption{Relative cost of each of the main services during periods of time when only one workflow was running at the ATLAS Google Site: MC Full Simulation, MC Reconstruction, Group Production, MC Event Generation and Data Reprocessing. The final
column shows the average values for the full duration of the project, from July 2022 to September 2023.}
\label{tab:singleWorkflowCosts}
{\footnotesize
\begin{tabular*}{\textwidth}{@{\extracolsep{\fill}}|l||l|l|l|l|l||l|}
\hline
& MC Full & MC & Group & MC Event & Data & Full \\
& Simulation & Reconstruction & Production & Generation & Reprocessing & Project \\
& 06/01--19/01 & 05/02--09/02 & 11/02--19/02 & 07/04--12/04 & 12/07--16/07 & 07/22--09/23 \\
\hline
Compute        & 73\% & 65\% & 52\% & 47\% & 21\% & 28\% \\
Storage        & 10\% & 10\% &  9\% & 32\% & 11\% & 20\% \\
Network egress &  7\% & 11\% & 26\% &  3\% & 63\% & 46\% \\
Other          & 10\% & 14\% & 14\% & 19\% &  4\% &  6\% \\
\hline
\end{tabular*}
}
\end{table}

User Analysis workflows have also been running on the ATLAS Google site since March 2023, as can be seen in Figure~\ref{fig:singleWorkflowsA}.
Whilst it would have been desirable to run only User Analysis at the ATLAS Google site for some period, this proved difficult, primarily due to the unpredictable nature of such workflows, compared to the rather standard production workflows employed in the single job type periods described in this section.
Despite replicating several popular analysis datasets to the site, it was difficult to get enough user jobs in the queue, and ultimately it was not possible to run only analysis workflows on the site, although this was almost achieved in the second half of March.

\subsection{Resource bursting}
\label{sec:bursting}

Cloud computing is intrinsically highly elastic in its nature, and offers the opportunity to acquire a significant number of additional resources, potentially at short notice.
In the context of Active Learning~\cite{ATL-PHYS-PUB-2022-045} this is particularly advantageous, when it is essential to increase the speed of each iteration of MC sample production.
Previous studies~\cite{chep23clouds} have shown that it is possible to quickly ramp up many tens of thousands of job slots at Google and process a small number of MC events through all steps in the MC production chain to arrive at the DAOD used for analysis.

Bursting a large amount of additional compute capacity may also be useful if for example a particular MC sample is urgently required, and this scenario was explored at the ATLAS Google site in June 2023.
In this case a 50M event standard top--quark pair production MC sample was chosen to undergo Full Simulation as quickly as possible, by bursting to 100,000 job slots.
The task was configured as standard 2000 event 8--core jobs, which each take on average between 6--8 hours, so that all 50M events should be processed within 24 hours.
The input EVNT data was replicated to the Google storage and the site was drained of all other running jobs before starting the test.
Whilst this was not strictly required, it allowed the burst of resources to be isolated, which was useful for monitoring purposes.

\begin{figure}
\centering
\sidesubfloat[]{\label{fig:burstA}
\includegraphics[width=0.96\textwidth]{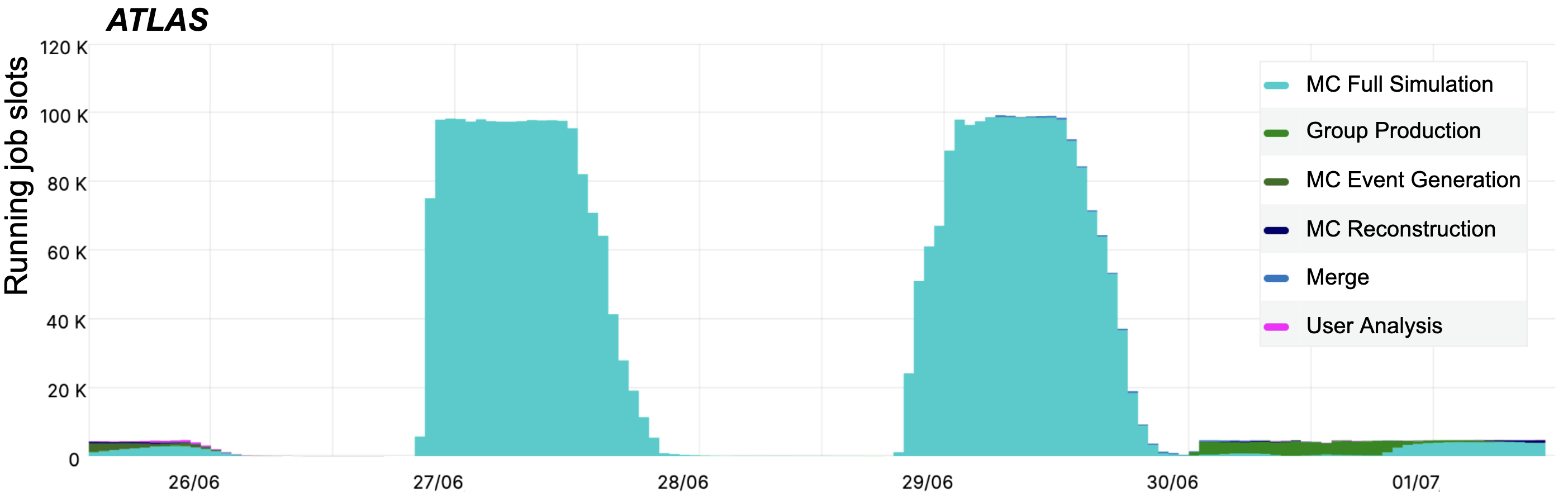}
}
\vspace{0.5cm}\\
\sidesubfloat[]{\label{fig:burstB}
\includegraphics[width=0.96\textwidth]{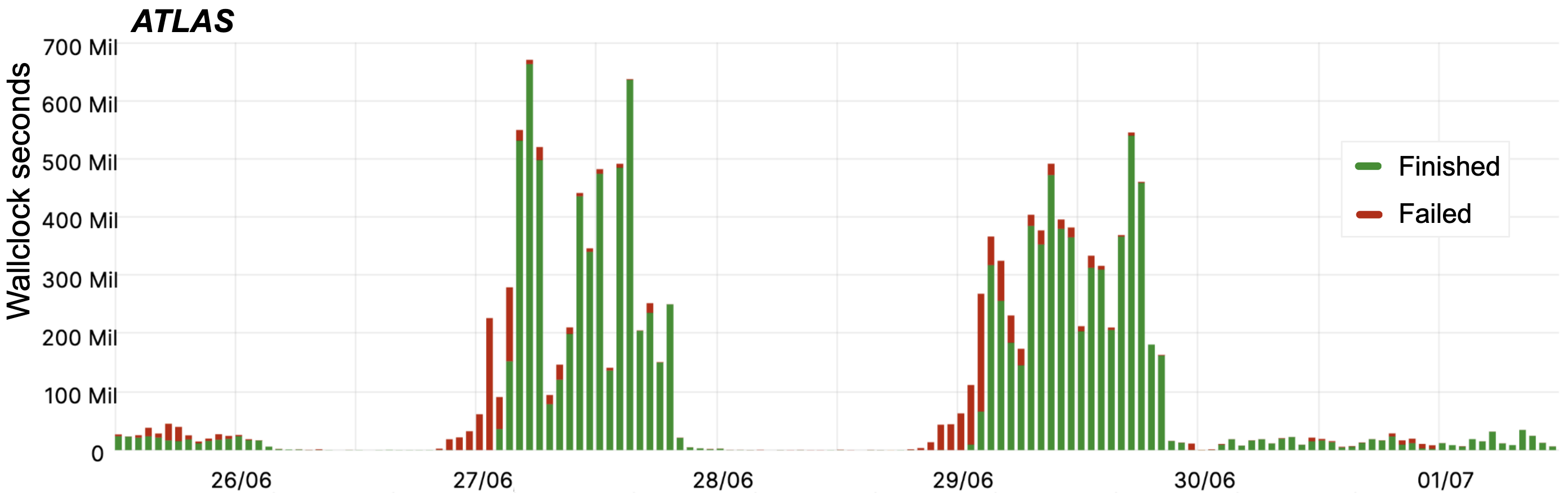}
}
\vspace{0.5cm}\\
\sidesubfloat[]{\label{fig:burstC}
\includegraphics[width=0.92\textwidth]{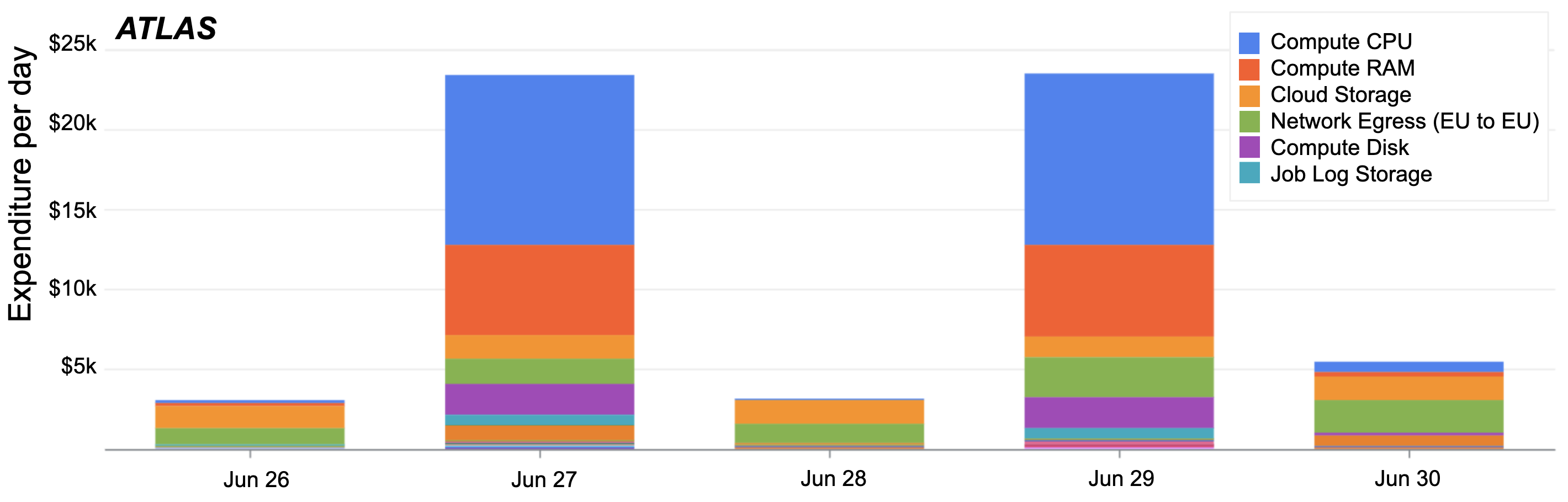}~~~
}
\caption{Distributions covering the resource burst tests done at the ATLAS Google site in June 2023. (a) The running jobs of the two bursts of MC Full Simulation. (b) The wall--clock consumption of the jobs running on the Google site. (c) The daily list--price cost per service from the Google billing console, where the compute contributions are seen to dominate on the burst days.}
\label{fig:burst}
\end{figure}

The results of the burst test are shown in Figure~\ref{fig:burst}. The ramp up to 100,000 running job slots was achieved in 1--2 hours without issue, as can be seen in Figure~\ref{fig:burstA}.
Wall--clock consumption is shown in Figure~\ref{fig:burstB}, which revealed a considerable amount of lost wall--time in the ramp up phase.
This was due to nodes accepting jobs while CVMFS was still being initialised, and hence the burst test was repeated with a slower ramp up profile.
In both cases, the overall lost wall--time was 11--13\%, somewhat more than observed on the grid, and coming from the ramp up phase and the low and constant level of preemptions throughout both tests.
Nevertheless, in both cases all 50M events were processed within 24 hours as expected.

The same MC Full Simulation sample has been processed several times on the various resources currently employed by ATLAS, without draining queues or using a dedicated site or queue.
Each of these tasks took between eight and ten days to process, even when a significant fraction of the work, between 30\% and 75\%, was executed on a few powerful sites such as the ATLAS High Level Trigger Farm~\cite{TRIG-2016-01,ATLAS-TRIG-2022-03} when used in Sim@P1~\cite{simp1,chep23simp1} configuration or the Vega~\cite{vega} and NERSC--Perlmutter~\cite{perlmutter} HPCs.
In this sense, the burst test can be considered a success, whilst at the same time having the advantage of exposing the cost of a well defined data processing activity.
The list--price cost of each burst run at the ATLAS Google site can be seen in Figure~\ref{fig:burstC}, where the sum of the compute based components is around \$23,000 each time.

\subsection{Scaling up the site again}
\label{sec:scaling-up-again}

\begin{figure}
\centering
\sidesubfloat[]{\label{fig:scalingUpAgainA}
\includegraphics[width=0.96\textwidth]{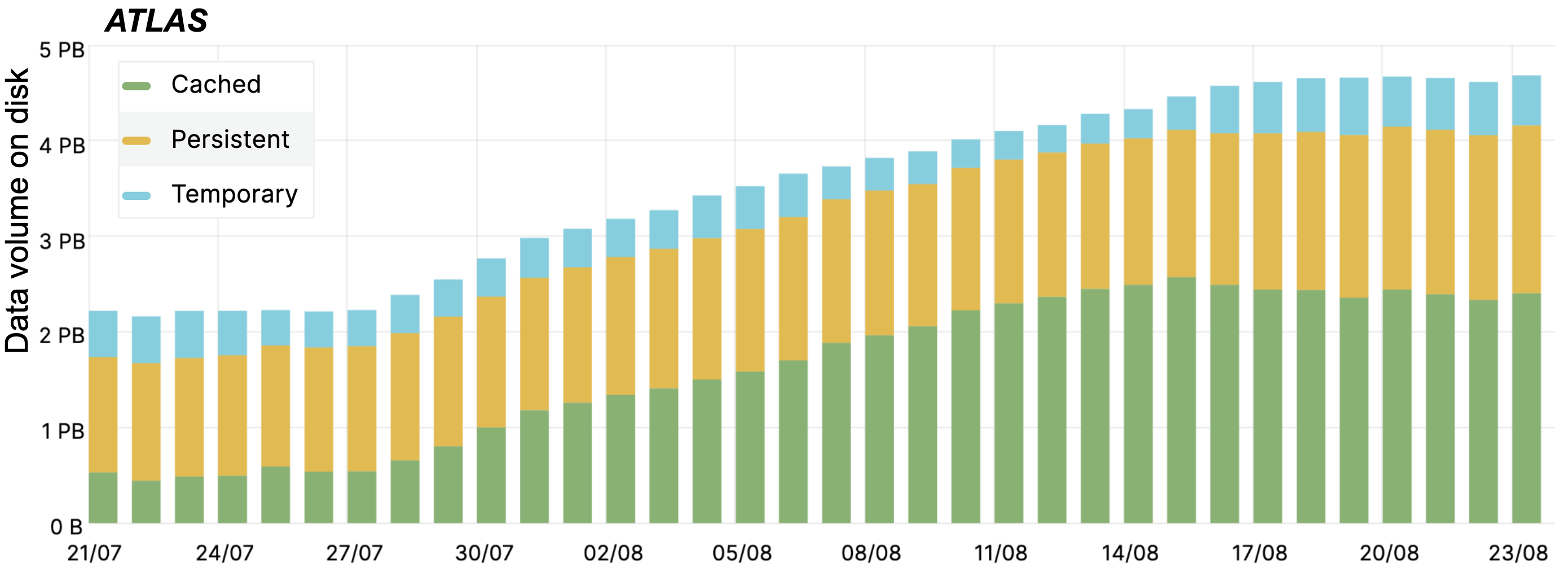}
}
\vspace{0.5cm}\\
\sidesubfloat[]{\label{fig:scalingUpAgainB}
\includegraphics[width=0.96\textwidth]{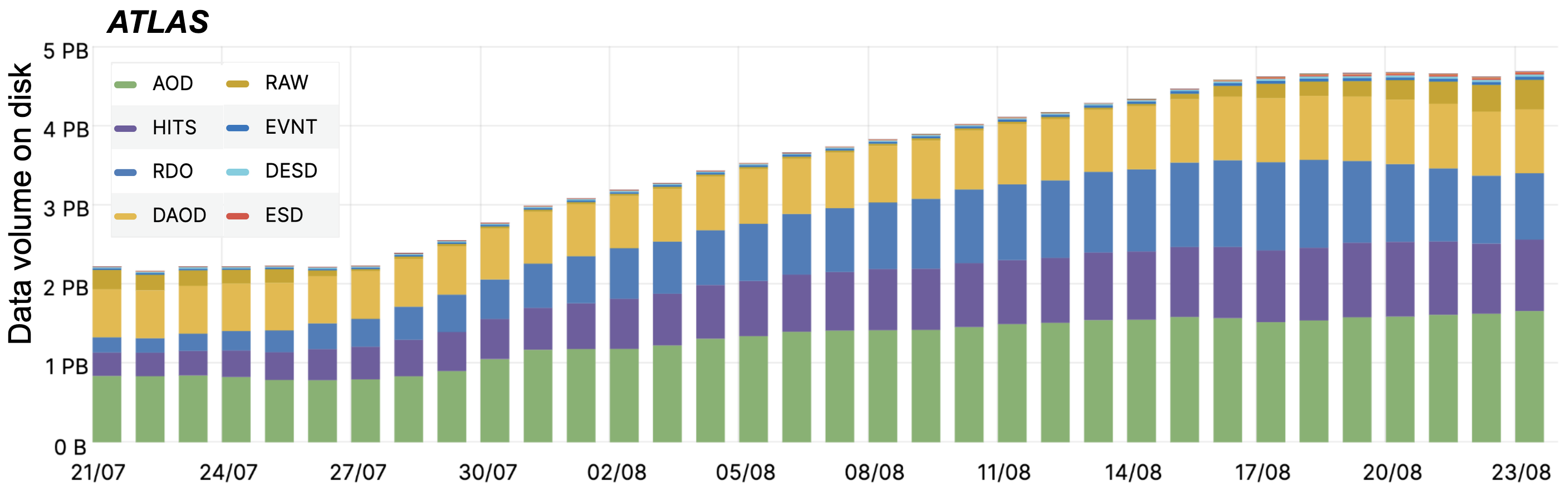}
}
\vspace{0.5cm}\\
\sidesubfloat[]{\label{fig:scalingUpAgainC}
\includegraphics[width=0.96\textwidth]{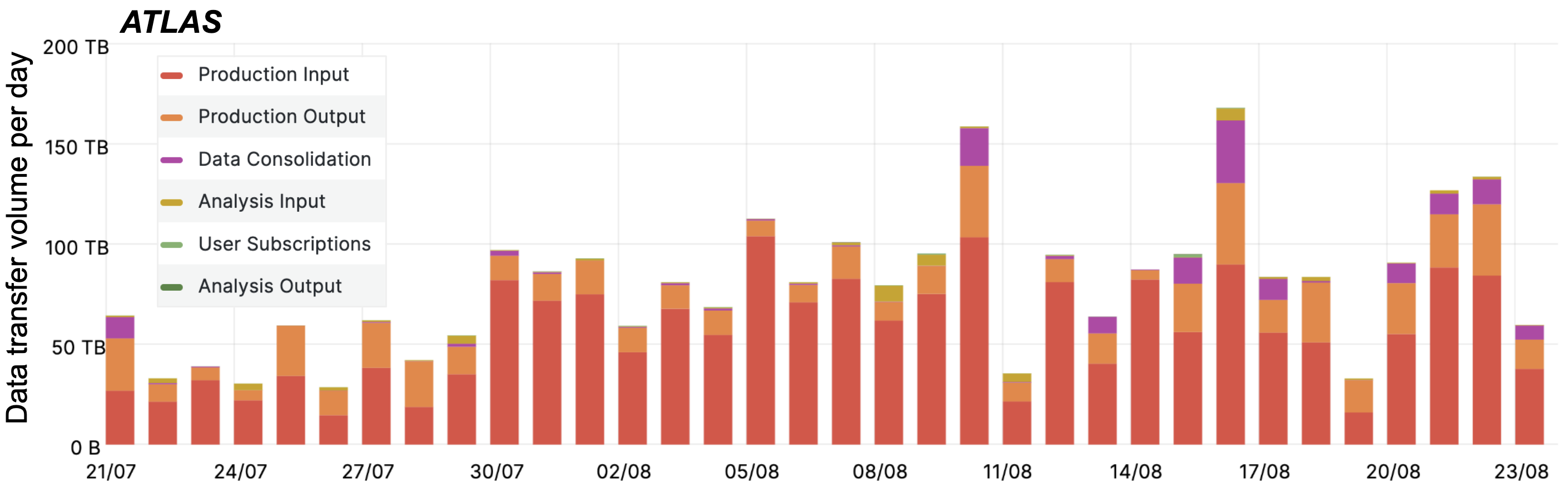}
}
\caption{Data stored at the ATLAS Google site: (a) grouped into different replica types and (b) grouped into different ATLAS data formats. (c) The different types of daily transfers out of the ATLAS Google site.}
\label{fig:scalingUpAgain}
\end{figure}

For the final few weeks of the current project, the size of the ATLAS Google site was scaled up again to between the size of an ATLAS Tier--1 and Tier--2 grid site, with around 5~PB of storage and 10,000 running jobs slots, utilised by all job types.
This was done in July 2023, ahead of launching the data reprocessing single workflow period described in Section~\ref{sec:different-workflows}, and this configuration remained until the end of the project in September.
The 15 largest Tier--2 grid sites and about half of the Tier--1 sites typically each provide at least this number of job slots to ATLAS computing.
Almost all of the Tier--1 sites and the ten largest Tier--2 sites have at least 5~PB allocated to their DATADISKs, so at the end of July the size of the ATLAS Google Site DATADISK was increased from 2.5~PB to 5~PB.
At this time, the site was also reconfigured as a nucleus, so that not only unique data could be stored there but the site could also act as the output destination for a task running jobs at all other ATLAS grid sites.
Furthermore, the large distances to other RSEs set in an earlier phase of the project were somewhat reduced,
in particular to the German Tier--1 and associated Tier--2 sites, which are geographically close to the ATLAS Google site in Belgium.
Figure~\ref{fig:scalingUpAgain} shows the changes to the data stored and the transfers out during the month following these site reconfigurations.

Figure~\ref{fig:scalingUpAgainA} shows the data stored at the ATLAS Google site, grouped into three different replica types: Persistent, which has a Rucio rule with no lifetime (typically data placed at the site); Temporary, which has a Rucio rule with a lifetime (typically data replicated to the site via PanDA for production jobs, with a lifetime of two weeks) and Cached, which is data with no current Rucio rule and therefore may be deleted at any time.
The main consequence of the site changes was, as expected, an increase in the Cached data as a result of the increased
space available for output data from production tasks running at the site.
The volume of Persistent and Temporary data remains roughly constant.

The variety of data types stored at the ATLAS Google site during this period can be seen in Figure~\ref{fig:scalingUpAgainB}, where a marked increase in RDO files is visible.
These data are used as input to MC Reconstruction tasks in combination with HITS, which is the output of MC Simulation.
It is likely that the nucleus nature of the site, in combination with the reduced distances, means that more HITS datasets remain on the RSE and are available as favourable input for MC Reconstruction tasks, both at the ATLAS Google site and elsewhere.
A steady increase in HITS and AOD (the output of MC Reconstruction) is also observed as well as the presence of some RAW data files used as input to data reprocessing tasks.

Figure~\ref{fig:scalingUpAgainC} shows the daily transfers out of the ATLAS Google site, which show a steady increase after the end of July.
Most of the egress is due to files replicated from the site to use as production input elsewhere, for example AODs to be used as input for the production of analysis level data (DAOD) at another ATLAS grid site.
The level of egress is however significantly less than in 2022, which at its peak had a rate of more than 1~PB per week (see Figure~\ref{fig:firstSixMonthsC}).
This reduction is likely due to two main reasons.
Firstly, compared to 2022, the short distances set in Rucio between RSEs were limited to only the sites in the German cloud.
Secondly, there was a higher fraction of new data at the Google RSE when it was first activated.
At the beginning of the project, the RSE was empty and all data were newly written there, whereas in July 2023 there was already around 2~PB stored there before the site was scaled up again.

Decommissioning of the ATLAS Google site took place in September 2023, so that all resources employed during the project were effectively switched off from the holistic perspective of ATLAS distributed computing.
The PanDA queue was disabled mid--September and all unique data moved to one of the ATLAS Tier--1 sites.
Any remaining user data from R\&D projects were removed and the site was fully decommissioned by September 21st.
No significant difficulties were encountered during this process, which was similar to the usual decommissioning of a typical ATLAS grid site.

\section{Cloud and network costs}
\label{sec:network-costs}

Deploying grid resources in commercial clouds creates a demand for networking services that can have an impact on performance and potentially also on cost.
Egress traffic is a particularly expensive resource in the cloud, due in part to the commercial strategy of providers to incentivise to continue to use their resources and not migrate to other cloud providers.
At the time of writing, the list--price for storing data in the Google \emph{europe--west1} region is \$20 per TB per month~\cite{googleStoragePricing}, while the price for egressing data is between \$45 and \$85 per TB, depending on the volume~\cite{googleNetworkPricing}.

\begin{figure}
\sidesubfloat[]{\label{fig:costBreakdownA}
\includegraphics[width=0.94\textwidth]{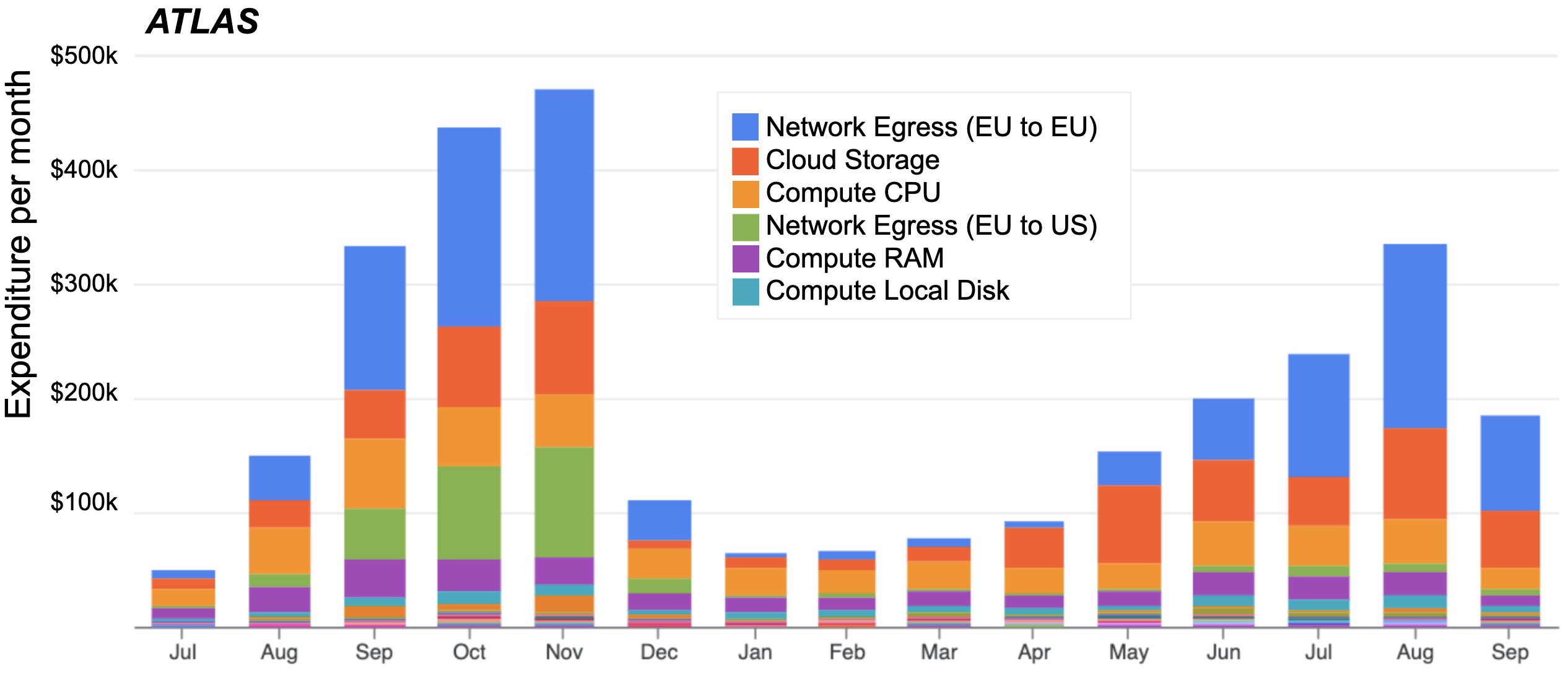}~~~~
}
\vspace{0.5cm}\\
\sidesubfloat[]{\label{fig:costBreakdownB}
\includegraphics[width=0.94\textwidth]{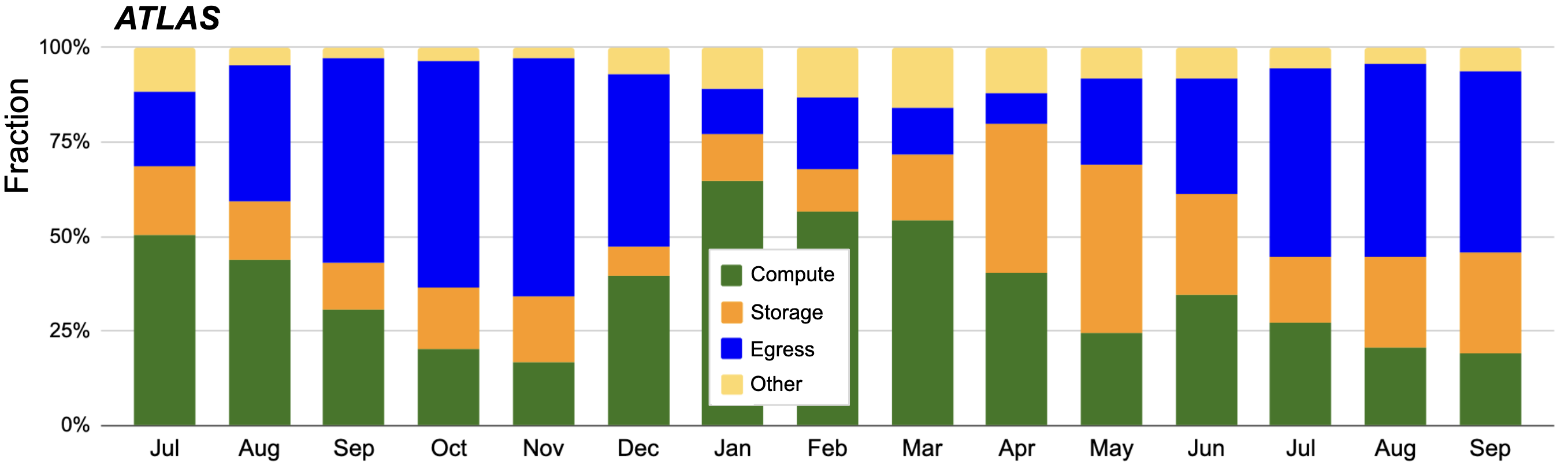}
}
\caption{Cost breakdown for the ATLAS Google site for the period July 2022 to September 2023. (a) The monthly list--price cost per service from the Google billing console. (b) The percentage contribution to the monthly cost from each of the services grouped as indicated in the legend.}
\label{fig:costBreakdown}
\end{figure}

As previously described, probably the most important feature of the cloud resources is that the costs involved are heavily dependent on which services are used.
This dependence can be seen in the cost breakdown of running the ATLAS Google site, which varied significantly month to month, or day to day, depending on the activity of the site.
Figure~\ref{fig:costBreakdownA} shows the monthly list--price cost profile for the full 15 month duration of the project.
The cost of compute is basically stable, only showing small variations at the times when the number of running job slots at the sites was changed (for example when the site was increased from five to ten thousand job slots from August to November 2022, and the CPU burst test in June 2023).
However, the cost of storage and network egress vary considerably, depending on the dominant activity.

Figure~\ref{fig:costBreakdownB} shows the relative fraction of each service to the total monthly cost.
For the first months of the project, until November 2022, the egress cost increased as new data accumulated at the site and jobs running at other grid sites accessed these data.
By November 2022, the costs associated with egress reached 54\% of the monthly total.
Other patterns can be observed in 2023, such as in April and May when the analysis input data was replicated to the ATLAS Google site, correspondingly increasing the fraction of the total cost spent on storage.
The increase in egress due to the data reprocessing performed at Google is also visible in July.

\subsection{Dedicated networks}
\label{sec:dedicated-networks}

Egress network traffic is the resource that can have the largest impact on the running costs over short timescales.
It is an expensive resource, and its use can increase very rapidly if some data at the site become popular and are suddenly accessed by thousands of jobs at other sites.
Moreover, the data going from the ATLAS Google site to other ATLAS grid sites do so through the general purpose internet, as opposed to using the LHCONE/LHCOPN private networks that link most of the ATLAS sites.
This has the effect of generating potentially very large traffic over the internet link into the destination sites, which can have cost implications at the destination, since some sites have higher costs or lower available bandwidth associated with their general purpose internet links compared to the dedicated LHCONE links.
The utilisation of such links can lead to operational disruptions, potentially impacting a site's availability to its users.
This is often due to lower provisioning of general internet bandwidth by the sites, primarily stemming from the associated higher costs, resulting in rapid saturation.
It is therefore important to investigate mechanisms that could control and potentially reduce the egress costs.
One such mechanism is dedicated network links with the cloud providers.

The price for sending traffic through a dedicated link~\cite{googleInterconnectPricing} has two components: a fixed one of the order of \$2.4 per hour for a 10~Gbps circuit for instance, plus a variable one that depends on traffic, priced at \$20 per TB.
Additional costs might also arise, depending on the specific service provider and route of the intermediate connection. According to this, routing the traffic exiting the ATLAS Google site through a dedicated link could potentially reduce the egress costs to less than half if traffic above 3~PB per month is generated.
A recent study within the IceCube Collaboration~\cite{Sfiligoi:2021jsc} came to a similar conclusion, where by employing dedicated network links the egress costs for data-intensive applications could be reduced by between 50\% and 75\%.
It is however worth emphasising that the scope of the study undertaken by IceCube significantly differs from that of ATLAS.
IceCube utilised dedicated links connecting the cloud to a specific site at UW-Madison, whereas ATLAS is currently investigating the feasibility of deploying dedicated links to establish connections between cloud resources and the LHCONE overlay network, facilitating data transfers to numerous sites worldwide.
It is therefore important to recognize that the complexity and potentially the cost associated with these two approaches may differ considerably.

To explore this option, an engagement has been started with ESnet~\cite{esnet} to provision a dedicated 10~Gbps link to the Google \emph{us--east4} cloud through their ESnet Cloud Connect service.
The aim of this exercise is to test two things: first, to measure and confirm a reduction of the egress costs for large data transfers, and second, to try to route the egress traffic from the ATLAS Google site into LHCONE to avoid downstream costs associated to high--volume traffic through the regular internet links at receiving sites.

The expectation is that dedicated network links will provide a way to lower the networking costs, but at the same time they will also add complexity to the deployment and operations of the cloud resources.
Moreover, besides the technical complexity of provisioning the dedicated network links, there is also an organisational complexity that arises from the fact that the implementation of this setup will vary depending on the combination of cloud provider (and even region inside a provider) and NREN that provides the peering.
In different countries, NRENs will have different capabilities and conditions for peering with cloud providers.
Furthermore each cloud provider likely offers their own specific tools to deploy and manage the dedicated links, each with very different provisioning procedures.
Another consideration would be if multiple experiments, for example ATLAS and CMS, were to provision resources from the same cloud provider whether they could both use the same dedicated network link.
There is clearly a significant programme of work in this area, beyond the time frame of the current AGP.

\section{Feedback from grid site administrators}
\label{sec:feedback}

As part of this study, feedback was collected from several system administrators representing various ATLAS Tier--1 grid sites.
In particular, their experiences with commercial clouds were discussed, to gain some insights into the primary advantages and drawbacks associated with these services.

\subsection{Concerns related to cloud computing and comparisons to this project}
\label{sec:concerns}

It was a common view among site administrators that cloud is more expensive than on--premises solutions.
In particular, concerns were raised about high egress costs, which can significantly impact the overall expenses.
Some of these comparisons may have been made against dedicated compute instances rather than spot instances, which have a higher cost.

Past experiences~\cite{Holzman2017} with spot instances revealed eviction rates of up to 15\%, which administrators considered unacceptably high.
This issue becomes particularly critical since many sites support multiple users beyond LHC activities, and these users might be less tolerant to evictions than the LHC experiments, affecting the overall service reliability.
During this project preemption was observed to be significantly lower, where only around 20\% of all failed jobs were due to evictions.
With an overall rate of 5\% lost wall--clock from failed jobs over the duration of the project, the resulting eviction rate of between 1--2\% compares favourably to the previous result described above.
However it is important to note that eviction rates for spot instances can vary based on several factors, including the time of the year, the cloud provider, the geographical region, the volume, and the types of resources utilised.

Worries were also expressed about unpredictable performance variations over time in cloud environments.
Administrators were concerned that cloud providers might change the underlying hardware behind a specific instance type, leading to potential performance fluctuations that could affect the quality of services.
The PanDA queue associated with the ATLAS Google site was very stable, experiencing negligible downtime throughout the duration of the project.
While some cloud providers do not specify the exact provided CPU models, within CPU families it is possible to define a preference for the newer generations.
The masked CPU values (publishing clock frequency and cache size) collected by the workload management system were homogeneous throughout the duration of the project, indicating a stable performance.
The adoption of the new, more flexible HEPScore~\cite{giordano2023hepscore} benchmarking model should provide further information about this topic in the near future.

Site administrators expressed a general concern about the risks associated with vendor lock--in.
They emphasised the importance of maintaining flexibility and the ability to migrate between different cloud providers to avoid being tied to a single vendor and potentially facing challenges with cost escalation, integration or data portability.
The solutions implemented by this project to interface Google Cloud resources are essentially cloud--agnostic, not only avoiding vendor lock--in but also potentially enabling access to other commercial cloud resources, as demonstrated at a lower scale at AWS~\cite{awsPricing}.

Data ownership and control were significant worries.
Another concern was about having critical data solely stored in the cloud, without having direct control over the physical infrastructure where the data is hosted.
Ensuring digital sovereignty is seen as crucial when handling data from unique scientific experiments.
The Google Cloud data access policy is clearly and strictly defined~\cite{googleAccess}, and user data is fully protected and accessible only to the customer cloud administrators, designated users, and contract managers.
Ultimately, ATLAS retains data (and algorithm) ownership when running in the Google Cloud.
In addition, an assurance on privacy is made that customer data will not be used for any commercial use or for training purposes.
A further, related point of concern is that as cloud resources are essentially leased, there is no opportunity to ensure that the hardware is responsibly and sustainably deployed to its maximum capability and longevity.

According to the feedback gathered from one site administrator, the cost breakdown of operating a grid site generally comprises approximately one--third for personnel, one--third for operational expenses, and one--third for hardware investments.
The technical personnel effort required to operate one of the ATLAS Tier--1 centres is estimated to be around 10 FTEs.
Some site administrators hold the view that even with a substantial migration of resources to the cloud, the essential operational effort needed to run a site would not experience a significant reduction.
This is because only a limited number of hardware--oriented technical positions may no longer be required.
However, some issues arise that make this comparison inherently challenging.
The experience of the Google site primarily focuses on the operational effort required to provide CPU and storage at scale for a
single experiment.
In contrast, the roles and responsibilities of technical personnel at on--premises sites may be much broader in scope.

Finally, site administrators emphasised that funding agencies are making substantial investments in building new, energy--efficient data centres, which does not indicate a trend or incentive to increase the utilisation of cloud resources.
In this context, Google cloud resources operate with net--zero operational greenhouse gas emissions by neutralising any remainder via investing in carbon offsets.
Google data centres have above average Power Usage Effectiveness~\cite{googleEfficiency} and provide dynamic information about the carbon intensity of their regions, enabling users to steer their load to minimise emissions.
In particular the site utilised by this project, \emph{europe--west1}, has one of the lowest grid carbon intensities~\cite{googleCarbon} among Google data centres.

\subsection{Purchasing procedures}
\label{sec:purchasing}

Public institutions, including those operating grid sites, often encounter the requirement to procure services through public tendering processes.
However, when it comes to large--scale purchasing of cloud services, the administrative demands involved can be quite difficult.
In response to this challenge, the OCRE~\cite{ocre} project was initiated in 2019 with the aim of streamlining the procurement process for cloud services in Europe.

For organisations contracting cloud services within Europe today, utilising a framework like OCRE becomes a viable option to do their purchasing.
OCRE facilitates this purchasing process through NRENs, which in turn maintain lists of country--specific authorised resellers that offer cloud services covered under the OCRE framework agreements.
One of the key benefits provided by OCRE is the provisioning of a standardised contract.
While this contract serves as a baseline, further negotiations at the country or site level are possible.
These negotiations could lead to larger volume discounts, ultimately benefiting the organisations involved.

Whilst OCRE is available for institutions within Europe, it is not applicable in countries other than the forty members of the framework.
The process for contracting cloud services in the USA, for example via HEPCloud~\cite{hepCloud} or CloudBank~\cite{cloudBank}, or in Asia may be completely different and may result in different pricing conditions compared to those in Europe.
It is crucial to recognise that the landscape for cloud service procurement can vary between regions, necessitating tailored approaches based on the specific regulatory and contractual requirements of each region.
If multiple sites were to buy into the same cloud provider, some of these differences could be overcome, although remaining administrative hurdles would need to be clarified.

For international organisations like ATLAS seeking to implement a coherent strategy for utilising cloud services, the varying regulations and procurement processes across regions add complexity to the management and planning efforts.
Flexibility and adaptability become essential for navigating the diverse cloud landscapes and effectively leveraging cloud resources across different geographical areas.

\section{Further investigations and future work}
\label{sec:future}

There are several areas that would benefit from an extension of the ATLAS Google Project, to allow further exploration of employing commercial cloud resources for ATLAS.

Firstly, this could involve further investigation of the ATLAS Google Site, evaluating the typical grid site configuration but at a lower level of resources than deployed at the end of this project cycle.
As discussed in Section~\ref{sec:scaling-up-again}, whilst the egress is more under control it is nevertheless still significant, so a more detailed evaluation of which workflows are suitable for commercial cloud should be done, including any necessary, additional configuration changes.
Understanding how the subscription agreement structure works and what a reasonable discount looks like is also an important factor when considering workflow restrictions, site structure adjustments or long--term contract costs.
The impact of these studies could also be expanded to potentially evolve the ATLAS workflow and data management systems to take into account the cost of the network.

Secondly, peering with commercial cloud networks would be the main focus for further investigation, as this multi--faceted, time--intensive activity is only just beginning now.
The network costs incurred due to transfers via the internet are a critical roadblock to widespread commercial cloud storage adoption for scientific computing overall and in the case of the LHC experiments, transfers need to be routed through the LHCONE overlay to eliminate these costs for the WLCG sites.
At the same time, data scheduling should be introduced to reduce the data volumes that incur transfer and storage costs.
Widespread peering is necessary to reduce the need to route via complicated paths.
There are several NRENs in the US and EU, which should be approached to discuss the technical details of network peering, for example ESnet~\cite{esnetPeering}, Internet2~\cite{internet2}, or G\'{E}ANT Network~\cite{geant}.

There are various options in the Google Network stack to support this activity, but at the same time the necessary Google documentation does not seem to be publicly available.
Detailed discussions would be needed between WLCG and CERN IT network experts together with the corresponding network experts from Google.
If necessary, short--term Google premium support could be financed once the peering options with the NRENs have been explored.

There is the potential need to develop new features in the Rucio~\cite{barisits2019rucio}, GFAL~\cite{gfal}, Davix~\cite{davix}, and FTS~\cite{fts} stack to support this R\&D and ATLAS would continue to work with the respective teams to discuss the objectives and milestones, and to follow the implementation, deployment, and operations.
There are two distinct analyses and respective evolutions of the ATLAS computing model that could be explored.
One idea is to improve the necessary workflow and data management policies when using cloud storage.
This would involve for example only storing data in the cloud that is currently in use, which is rather different to the current grid storage model.
In this way, the complexity of a heterogeneous infrastructure setup reduces the egress volume, but still allows bursting to cloud compute with large data inputs.
Alternatively, the necessary capabilities to exploit cloud storage and network features such as bucket--level copy could be developed, to facilitate internal transfers between different cloud regions.
This would remove the need to egress the data via FTS, which incurs the usual associated transfer costs.

Thirdly, beyond the ATLAS Google Site configuration and network considerations, many areas of R\&D were performed as part of the AGP, using multiple services offered within the Google Cloud~\cite{megino2024operational}.
Most of these projects have taken advantage of the elastic availability of special types of resources to ramp up and down ephemeral compute clusters using for example GPUs, large amounts of memory, or ARM CPUs depending on the current need.
This was done either through the ATLAS PanDA workflow management system or using interactive compute with Jupyter~\cite{jupyterHub} notebooks and Dask~\cite{dask} task scheduling.

The usage of such non--standard resources that are not easily available at standard WLCG grid sites has proven to be extremely valuable and effective, helping to develop and expedite new ATLAS data analysis techniques using machine learning and the migration of the ATLAS software to ARM CPUs.
Compact data formats with columnar data access have been investigated using Google resources.

ATLAS plans to continue to take advantage of non--standard resources like GPUs and ARM CPUs to work on innovative and novel analysis and software techniques, accelerating the process to discoveries.
Another larger focus area could be the development of high energy phsyics algorithms using tools not available before, such as Large Language Models and Generative Artificial Intelligence.
A continuation of the AGP will facilitate these initiatives, providing proven access to such resources.
Like the other activities proposed in this section they are not expected to be resource or cost intensive.

There are several other topics of interest that might be investigated during a continuation of the AGP.
Running analysis jobs at the ATLAS Google Site was only briefly examined and this workflow, which is unique in its highly variable nature, may warrant further scrutiny.
In particular, actions to replicate the most highly requested analysis level data samples were not measurably successful, requiring further understanding of data popularity and data placement.
Taking running analysis on the Google Cloud further, the idea of extending the resources of the ATLAS Google site with user--specific Google credits has also been raised.
Authentication issues could also be interesting to look into, when considering using the site not only for ATLAS data but also for hosting Open Data, available to all.
On the other hand, it may be worth investing some time in understanding the privacy implications of the data stored in commercial cloud resources, which may contain protected information, at least according to GDPR~\cite{gdpr}, and whether this must be taken into consideration.

The ATLAS Google Project has provided valuable insights into the use of commercial clouds, and there remain many avenues of investigation that could be pursued.
This next phase could feature a significant reduction of resources required, and hence financial expense, as the focus shifts to network connectivity and continued R\&D.
The ATLAS Google site could nevertheless still continue, albeit at a lower level, although if the network R\&D is successful in reducing the egress cost, it could once again be ramped up to verify the savings arising from the implementation of dedicated peering solutions.

\section{Summary}
\label{sec:summary}

While traditional, site--based resources have always formed the backbone of ATLAS computing, commercial clouds may in some cases provide a viable and attractive alternative or addition.
Much experience was gained in the integration of the ATLAS Google site into ATLAS distributed computing and no significant technical issue was discovered to prevent the experiment employing such resources in the future.
Furthermore, the current workflow and data management tools employed by ATLAS are shown to be adequate for applying changes to the site configuration.
The technical solutions implemented are essentially cloud--agnostic, not only avoiding vendor lock--in but also potentially enabling access to other commercial cloud resources.
The subscription pricing model applied in this project has proven to be beneficial to ATLAS, although questions remain how this may change going forward.

The project has shown that commercial cloud sites are an effective mechanism for providing additional, on demand CPU resources.
At the level employed by ATLAS, typically five or ten thousand cores, preemption of the allocated job slots is barely an issue, even when using the spot instance model as is done here.
Whilst higher eviction rates were occasionally noticeable, for example during the data reprocessing single workflow period, the overall failure rate was not significantly higher than that observed on the grid.
The ATLAS Google site was also shown to be extremely effective as a bursting resource, quickly providing up to one hundred thousand additional job slots, resulting in a significantly faster production turnaround than is possible on the ATLAS grid sites.
In addition, the project has enabled parallel R\&D efforts to flourish by providing different types of resources, for example GPU or ARM, on an elastic basis, demonstrating rapid integration~\cite{megino2024operational}.

Whilst it was also shown that it is possible to integrate, adjust and expand associated storage at the cloud site, this is less trivial than CPU as the intrinsic network costs must be taken into consideration.
Storage and in particular network costs are known to dominate the TCO of commercial clouds, so much so this often dissuades sites taking an active interest in employing such resources.
The studies performed during the AGP and outlined in this TCO analysis have shown that commercial cloud is a technically viable option for ATLAS distributed computing, albeit with additional costs not necessarily considered when employing traditional grid resources.
Within the WLCG model the cost of the network is sometimes hidden, and whilst in reality this is probably rather high and means to reduce it are worth investigating, it can also be considered as irreducible and somewhat independent of ATLAS.
Conversely, commercial cloud data transfers over standard internet networks incur significant costs for the data centres.

The TCO evaluation has shown that without the subscription model, the cost of commercial cloud resources is significantly increased.
The Google Cloud resources used during this project cost a total of \$3.162M at list--price compared to the \$849,458 paid via the subscription agreement, representing a discount of 73\%.
Alternatively, ATLAS used 3.72 times more Google Cloud resources than were purchased via the subscription agreement, which means the resources used during this project would have been 272\% more expensive at list--price.
This is most obvious in the costs associated with the bursting test shown in Figure~\ref{fig:burstC}, which depicts daily expenditure considerably in excess of the \$1900 per day rate of the subscription agreement.

As shown in Table~\ref{tab:singleWorkflowCosts}, almost half of the total list--price costs are due to egress.
With this in mind, the ATLAS Collaboration is investigating ways to reduce this cost via dedicated network solutions, as outlined in Section~\ref{sec:dedicated-networks}.
It was also shown that egress costs are workflow dependent, which may be a consideration when employing such resources in the future.
In particular, the substantial egress associated with data reprocessing means that for now this workflow is best avoided until further improvements in network connectivity are deployed, and that until then the site cannot be seen as universally suitable for ATLAS.
If workflows such as Fast or Full Chain~\cite{fastChain} can be employed, where the egress of intermediate MC formats is avoided, this will also help to reduce these costs.

Establishing a viable and cost effective subscription agreement between experiment and commercial cloud provider is clearly a critical consideration of the TCO, given the large discrepancy between the list--prices and the agreement associated with this project.
A collaborative approach may be necessary to obtain the best deal with a large volume discount.
One option could be for CERN to make a significant purchase of cloud capacity and give the option to pay to be part of it, which has been done before~\cite{cloudbankEUNGI}.
This offer could extend not only to multiple sites, but also multiple experiments, and could even be done cooperatively with other international organisations such as EMBL~\cite{embl}.
Another important consideration going forward, especially if commercial clouds are employed by a significant number of sites, is to understand how such resources fit into the WLCG pledge structure.

The initial concerns about a significant loss of on--site personnel when outsourcing computing to the cloud, and the potential wider implications for ATLAS due to additional parallel support roles, appear to be less pronounced.
This is because, for the most part, these individuals at the sites would still be needed to contribute to ATLAS distributed computing.
As such, at least for larger sites, employing commercial clouds and off--premises resources may not immediately result in significant cost savings.

In summary, commercial cloud computing is an effective technical solution for ATLAS for providing additional CPU resources, and whilst the seamless integration of cloud--based storage was also achieved, network costs may be significant, based on the list--price.
Some ATLAS workflows are found to be better than others with respect to egress. Resource bursting was shown to be very effective, albeit at significant cost.
Establishing a favourable subscription agreement model that makes sense to both the cloud provider and the client is an advantage.
By leveraging the Google Cloud Subscription Agreement pricing model, ATLAS has effectively harnessed between three to four times the resources compared to what the same investment would deliver for the list--price.
It is yet to be seen how much this project influences the structure and cost of any potential follow up deal to be brokered.
There is much interest for ATLAS to continue this project, where network connectivity would be the main focus.

\section*{Acknowledgements}
%

% The next lines are included from the .//acknowledgements/Acknowledgements.tex input file
%
%

%
%

We thank CERN for the very successful operation of the LHC and its injectors, as well as the support staff at
CERN and at our institutions worldwide without whom ATLAS could not be operated efficiently.

The crucial computing support from all WLCG partners is acknowledged gratefully, in particular from CERN, the ATLAS Tier-1 facilities at TRIUMF/SFU (Canada), NDGF (Denmark, Norway, Sweden), CC-IN2P3 (France), KIT/GridKA (Germany), INFN-CNAF (Italy), NL-T1 (Netherlands), PIC (Spain), RAL (UK) and BNL (USA), the Tier-2 facilities worldwide and large non-WLCG resource providers. Major contributors of computing resources are listed in Ref.~\cite{ATL-SOFT-PUB-2023-001}.

We gratefully acknowledge the support of ANPCyT, Argentina; YerPhI, Armenia; ARC, Australia; BMWFW and FWF, Austria; ANAS, Azerbaijan; CNPq and FAPESP, Brazil; NSERC, NRC and CFI, Canada; CERN; ANID, Chile; CAS, MOST and NSFC, China; Minciencias, Colombia; MEYS CR, Czech Republic; DNRF and DNSRC, Denmark; IN2P3-CNRS and CEA-DRF/IRFU, France; SRNSFG, Georgia; BMBF, HGF and MPG, Germany; GSRI, Greece; RGC and Hong Kong SAR, China; ISF and Benoziyo Center, Israel; INFN, Italy; MEXT and JSPS, Japan; CNRST, Morocco; NWO, Netherlands; RCN, Norway; MNiSW, Poland; FCT, Portugal; MNE/IFA, Romania; MESTD, Serbia; MSSR, Slovakia; ARRS and MIZ\v{S}, Slovenia; DSI/NRF, South Africa; MICINN, Spain; SRC and Wallenberg Foundation, Sweden; SERI, SNSF and Cantons of Bern and Geneva, Switzerland; MOST, Taipei; TENMAK, T\"urkiye; STFC, United Kingdom; DOE and NSF, United States of America.

Individual groups and members have received support from BCKDF, CANARIE, CRC and DRAC, Canada; CERN-CZ, FORTE and PRIMUS, Czech Republic; COST, ERC, ERDF, Horizon 2020, ICSC-NextGenerationEU and Marie Sk{\l}odowska-Curie Actions, European Union; Investissements d'Avenir Labex, Investissements d'Avenir Idex and ANR, France; DFG and AvH Foundation, Germany; Herakleitos, Thales and Aristeia programmes co-financed by EU-ESF and the Greek NSRF, Greece; BSF-NSF and MINERVA, Israel; NCN and NAWA, Poland; La Caixa Banking Foundation, CERCA Programme Generalitat de Catalunya and PROMETEO and GenT Programmes Generalitat Valenciana, Spain; G\"{o}ran Gustafssons Stiftelse, Sweden; The Royal Society and Leverhulme Trust, United Kingdom.

In addition, individual members wish to acknowledge support from Armenia: Yerevan Physics Institute (FAPERJ); CERN: European Organization for Nuclear Research (CERN PJAS); Chile: Agencia Nacional de Investigaci\'on y Desarrollo (FONDECYT 1230812, FONDECYT 1230987, FONDECYT 1240864); China: Chinese Ministry of Science and Technology (MOST-2023YFA1605700), National Natural Science Foundation of China (NSFC - 12175119, NSFC 12275265, NSFC-12075060); Czech Republic: Czech Science Foundation (GACR - 24-11373S), Ministry of Education Youth and Sports (FORTE CZ.02.01.01/00/22\_008/0004632), PRIMUS Research Programme (PRIMUS/21/SCI/017); EU: H2020 European Research Council (ERC - 101002463); European Union: European Research Council (ERC - 948254, ERC 101089007), Horizon 2020 Framework Programme (MUCCA - CHIST-ERA-19-XAI-00), European Union, Future Artificial Intelligence Research (FAIR-NextGenerationEU PE00000013), Italian Center for High Performance Computing, Big Data and Quantum Computing (ICSC, NextGenerationEU); France: Agence Nationale de la Recherche (ANR-20-CE31-0013, ANR-21-CE31-0013, ANR-21-CE31-0022), Investissements d'Avenir Labex (ANR-11-LABX-0012); Germany: Baden-Württemberg Stiftung (BW Stiftung-Postdoc Eliteprogramme), Deutsche Forschungsgemeinschaft (DFG - 469666862, DFG - CR 312/5-2); Italy: Istituto Nazionale di Fisica Nucleare (ICSC, NextGenerationEU); Japan: Japan Society for the Promotion of Science (JSPS KAKENHI JP22H01227, JSPS KAKENHI JP22H04944, JSPS KAKENHI JP22KK0227, JSPS KAKENHI JP23KK0245); Netherlands: Netherlands Organisation for Scientific Research (NWO Veni 2020 - VI.Veni.202.179); Norway: Research Council of Norway (RCN-314472); Poland: Polish National Agency for Academic Exchange (PPN/PPO/2020/1/00002/U/00001), Polish National Science Centre (NCN 2021/42/E/ST2/00350, NCN OPUS nr 2022/47/B/ST2/03059, NCN UMO-2019/34/E/ST2/00393, UMO-2020/37/B/ST2/01043, UMO-2021/40/C/ST2/00187, UMO-2022/47/O/ST2/00148, UMO-2023/49/B/ST2/04085); Slovenia: Slovenian Research Agency (ARIS grant J1-3010); Spain: Generalitat Valenciana (Artemisa, FEDER, IDIFEDER/2018/048), Ministry of Science and Innovation (MCIN \& NextGenEU PCI2022-135018-2, MICIN \& FEDER PID2021-125273NB, RYC2019-028510-I, RYC2020-030254-I, RYC2021-031273-I, RYC2022-038164-I), PROMETEO and GenT Programmes Generalitat Valenciana (CIDEGENT/2019/027); Sweden: Swedish Research Council (Swedish Research Council 2023-04654, VR 2018-00482, VR 2022-03845, VR 2022-04683, VR 2023-03403, VR grant 2021-03651), Knut and Alice Wallenberg Foundation (KAW 2018.0157, KAW 2018.0458, KAW 2019.0447, KAW 2022.0358); Switzerland: Swiss National Science Foundation (SNSF - PCEFP2\_194658); United Kingdom: Leverhulme Trust (Leverhulme Trust RPG-2020-004), Royal Society (NIF-R1-231091); United States of America: U.S. Department of Energy (ECA DE-AC02-76SF00515), Neubauer Family Foundation.

%
%

% End of text imported from the .//acknowledgements/Acknowledgements.tex input file

%
%
%
%
%
%
%

%
%
%
%
%
%

%
%

%
%
%
\printbibliography
\clearpage
% ATLAS Collaboration author list
% Reference date of SOFT-2023-02 is 2024-04-04
% Author list last updated on date 15-JUL-24
% Data extracted on 15-Jul-2024 for paper reference SOFT-2023-02
% at 4:20pm
 
\begin{flushleft}
\hypersetup{urlcolor=black}
{\Large The ATLAS Collaboration}

\bigskip

\AtlasOrcid[0000-0002-6665-4934]{G.~Aad}$^\textrm{\scriptsize 104}$,
\AtlasOrcid[0000-0001-7616-1554]{E.~Aakvaag}$^\textrm{\scriptsize 17}$,
\AtlasOrcid[0000-0002-5888-2734]{B.~Abbott}$^\textrm{\scriptsize 123}$,
\AtlasOrcid[0000-0002-0287-5869]{S.~Abdelhameed}$^\textrm{\scriptsize 119a}$,
\AtlasOrcid[0000-0002-1002-1652]{K.~Abeling}$^\textrm{\scriptsize 56}$,
\AtlasOrcid[0000-0001-5763-2760]{N.J.~Abicht}$^\textrm{\scriptsize 50}$,
\AtlasOrcid[0000-0002-8496-9294]{S.H.~Abidi}$^\textrm{\scriptsize 30}$,
\AtlasOrcid[0009-0003-6578-220X]{M.~Aboelela}$^\textrm{\scriptsize 45}$,
\AtlasOrcid[0000-0002-9987-2292]{A.~Aboulhorma}$^\textrm{\scriptsize 36e}$,
\AtlasOrcid[0000-0001-5329-6640]{H.~Abramowicz}$^\textrm{\scriptsize 154}$
% ...
\bigskip
\\

% ...
$^{17}$Department for Physics and Technology, University of Bergen, Bergen; Norway.\\
% ...
$^{30}$Physics Department, Brookhaven National Laboratory, Upton NY; United States of America.\\
% ...
$^{36}$$^{(a)}$Facult\'e des Sciences Ain Chock, R\'eseau Universitaire de Physique des Hautes Energies - Universit\'e Hassan II, Casablanca;$^{(b)}$Facult\'{e} des Sciences, Universit\'{e} Ibn-Tofail, K\'{e}nitra;$^{(c)}$Facult\'e des Sciences Semlalia, Universit\'e Cadi Ayyad, LPHEA-Marrakech;$^{(d)}$LPMR, Facult\'e des Sciences, Universit\'e Mohamed Premier, Oujda;$^{(e)}$Facult\'e des sciences, Universit\'e Mohammed V, Rabat;$^{(f)}$Institute of Applied Physics, Mohammed VI Polytechnic University, Ben Guerir; Morocco.\\
% ...
$^{45}$Physics Department, Southern Methodist University, Dallas TX; United States of America.\\
% ...
$^{50}$Fakult\"{a}t Physik , Technische Universit{\"a}t Dortmund, Dortmund; Germany.\\
% ...
$^{56}$II. Physikalisches Institut, Georg-August-Universit\"{a}t G\"ottingen, G\"ottingen; Germany.\\
% ...
$^{104}$CPPM, Aix-Marseille Universit\'e, CNRS/IN2P3, Marseille; France.\\
% ...
$^{119}$$^{(a)}$New York University Abu Dhabi, Abu Dhabi;$^{(b)}$United Arab Emirates University, Al Ain; United Arab Emirates.\\
% ...
$^{123}$Homer L. Dodge Department of Physics and Astronomy, University of Oklahoma, Norman OK; United States of America.\\
% ...
$^{154}$Raymond and Beverly Sackler School of Physics and Astronomy, Tel Aviv University, Tel Aviv; Israel.\\
% ...
$^{a}$ Also Affiliated with an institute covered by a cooperation agreement with CERN.\\
$^{b}$ Also at An-Najah National University, Nablus; Palestine.\\
$^{c}$ Also at Borough of Manhattan Community College, City University of New York, New York NY; United States of America.\\
$^{d}$ Also at Center for Interdisciplinary Research and Innovation (CIRI-AUTH), Thessaloniki; Greece.\\
$^{e}$ Associated at Centro de Investigaciones Energéticas, Medioambientales y Tecnológicas; Spain.\\
$^{f}$ Also at Centro Studi e Ricerche Enrico Fermi; Italy.\\
% ...

\end{flushleft}

% Created with Glance <Atlas.Glance@cern.ch>

%

%
%
%
%

%
%
%
%

\end{document}